\documentclass[preprint,12pt]{aastex}

\newcommand{\degree}{\ensuremath{^\circ}}

\begin{document}
\title{A Search for Dark Matter Annihilation \\
with the Whipple 10m Telescope}
\author{
  M. Wood\altaffilmark{1}, 
  G. Blaylock\altaffilmark{2},
  S. M. Bradbury\altaffilmark{3},
  J. H. Buckley\altaffilmark{4},
  K. L. Byrum\altaffilmark{5},
  Y. C. K. Chow\altaffilmark{1}, 
  W. Cui\altaffilmark{6},
  I. de la Calle Perez\altaffilmark{7},
  A. D. Falcone\altaffilmark{8},
  S. J. Fegan\altaffilmark{1},
  J. P. Finley\altaffilmark{6},
  J. Grube\altaffilmark{3},
  J. Hall\altaffilmark{9},
  D. Hanna\altaffilmark{10}, 
  J. Holder\altaffilmark{11}, 
  D. Horan\altaffilmark{5},
  T. B. Humensky\altaffilmark{12},
  D. B. Kieda\altaffilmark{13},
  J. Kildea\altaffilmark{14}, 
  A. Konopelko\altaffilmark{6}, 
  H. Krawczynski\altaffilmark{4},
  F. Krennrich\altaffilmark{15},
  M. J. Lang\altaffilmark{16},
  S. LeBohec\altaffilmark{13},
  T. Nagai\altaffilmark{15}, 
  R. A. Ong\altaffilmark{1}, 
  J. S. Perkins\altaffilmark{14},
  M. Pohl\altaffilmark{15}, 
  J. Quinn\altaffilmark{17},
  H. J. Rose\altaffilmark{3},
  G. H. Sembroski\altaffilmark{6}, 
  V. V. Vassiliev\altaffilmark{1},
  R. G. Wagner\altaffilmark{5},
  S. P. Wakely\altaffilmark{12},
  T. C. Weekes\altaffilmark{14},
  and A. Weinstein\altaffilmark{1}
}

\altaffiltext{1}{Department of Physics and Astronomy, University of
  California, Los Angeles, CA 90095, USA}
\altaffiltext{2}{Department of Physics, University of Massachussetts, 
  Amherst, MA 01003-4525, USA}
\altaffiltext{3}{School of Physics and Astronomy, University of Leeds, 
  Leeds, LS2 9JT, UK}
\altaffiltext{4}{Department of Physics, Washington University, St. Louis, 
  MO 63130, USA}
\altaffiltext{5}{Argonne National Laboratory, 9700 S. Cass Avenue, 
  Argonne, IL 60439, USA}
\altaffiltext{6}{Department of Physics, Purdue University, West Lafayette, 
  IN 47907, USA}
\altaffiltext{7}{Department of Physics, University of Oxford, 
  Oxford, OX1 3RH, UK}
\altaffiltext{8}{Department of Astronomy and Astrophysics, 525 
  Davey Lab.,Penn. State University, University Park, PA 16802, USA}
\altaffiltext{9}{Fermi National Accelerator Laboratory, Batavia, IL 60510, USA}
\altaffiltext{10}{Physics Department, McGill University, Montreal, 
  QC H3A 2T8, Canada}
\altaffiltext{11}{Department of Physics and Astronomy, University of Delaware,
Sharp Laboratory, Newark, DE 19716, USA}
\altaffiltext{12}{Enrico Fermi Institute, University of Chicago, 
  Chicago, IL 60637, USA}
\altaffiltext{13}{Physics Department, University of Utah, Salt Lake City, 
  UT 84112, USA}
\altaffiltext{14}{Fred Lawrence Whipple Observatory, Harvard-Smithsonian 
  Center for Astrophysics, Amado, AZ 85645, USA}
\altaffiltext{15}{Department of Physics and Astronomy, Iowa State University, 
  Ames, IA 50011, USA}
\altaffiltext{16}{Physics Department, National University of Ireland, 
  Galway, Ireland}
\altaffiltext{17}{School of Physics, University College Dublin, 
  Belfield, Dublin 4, Ireland}

\begin{abstract}
  We present observations of the dwarf galaxies Draco and Ursa Minor,
  the local group galaxies M32 and M33, and the globular cluster M15
  conducted with the Whipple 10m gamma-ray telescope to search for the
  gamma-ray signature of self-annihilating weakly interacting massive
  particles (WIMPs) which may constitute astrophysical dark matter
  (DM).  We review the motivations for selecting these sources based
  on their unique astrophysical environments and report the results of
  the data analysis which produced upper limits on excess rate of
  gamma rays for each source.  We consider models for the DM
  distribution in each source based on the available observational
  constraints and discuss possible scenarios for the enhancement of
  the gamma-ray luminosity.  Limits on the thermally averaged product
  of the total self-annihilation cross section and velocity of the
  WIMP, $\left<{\sigma}v\right>$, are derived using conservative
  estimates for the magnitude of the astrophysical contribution to the
  gamma-ray flux.  Although these limits do not constrain predictions
  from the currently favored theoretical models of supersymmetry
  (SUSY), future observations with VERITAS will probe a larger region
  of the WIMP parameter phase space, $\left<{\sigma}v\right>$ and WIMP
  particle mass ($m_\chi$).
\end{abstract}

\email{mdwood@astro.ucla.edu\\vvv@astro.ucla.edu}

\section{Introduction}

The existence of dark matter (DM) is supported by a variety of
observational data including measurements of the cosmic microwave
background \citep{spergel2007}, the large-scale distribution of
galaxies \citep{tegmark2004}, and gravitational lensing
\citep{clowe2006}.  In the $\Lambda$CDM cosmological model that is
currently favored by these data, DM comprises approximately $\sim$26\%
of the total energy density of the universe \citep{spergel2007}.
However, the nature of the particles that constitute DM remains
unknown.  A popular DM candidate is weakly interacting massive
particles (WIMPs), which existed in thermal equilibrium during the
early universe and later decoupled as the universe expanded.  Since
the time of decoupling, the WIMPs have remained non-relativistic,
behaving as a collisionless fluid on all but perhaps the shortest
spatial scales.  In order to reproduce the observed relic density of
DM, this hypothetical particle would need to have a cross section on
the scale of weak interactions.  A stable particle with these
properties, the lightest neutralino $\chi$, can be accommodated in
theories of supersymmetry (SUSY).

The mass of the neutralino is constrained to be $\gtrsim$ 6 GeV by CMB
measurements and accelerator searches \citep{bottino2003} and
$\lesssim$ 100 TeV by the unitarity limit on the thermal relic
\citep{griest1990}.  In the conventional SUSY scenarios, the
neutralino is a Majorana particle which can efficiently
self-annihilate in astrophysical environments with high DM density
producing secondary particles including high-energy gamma rays.  The
former and current generation of Air-Cherenkov Telescopes (ACTs),
including Whipple, HEGRA, CANGAROO-III, VERITAS, H.E.S.S., and MAGIC
are sensitive in the gamma-ray energy range from below 100 GeV to
above 10 TeV and can therefore make a substantial contribution to the
search for the signatures of DM self-annihilation.  Recently several
ACTs have detected gamma rays from the Galactic Center (G.C.)
\citep{albert2006,aharonian2004,kosack2004}.  Although a more
traditional astrophysical origin for this signal is currently favored
\citep{atoyan2004,aharonian2005}, DM self-annihilation has been
proposed as a possible explanation for these observations
\citep{horns2005}.

We present observations taken with the Whipple 10m telescope of five
astrophysical sources with the purpose of detecting the signature of
DM self-annihilation.  Section \ref{sourceReviewSection} summarizes
the motivations for selecting each source.  In Section
\ref{dmFluxSection} we discuss the signature of DM self-annihilation
into gamma rays and its dependence on the source astrophysics and the
particle physics properties of the WIMP.  In Section \ref{dataSection}
we review the atmospheric Cherenkov technique and the methods used to
analyze the data.  Results of the data analysis are described in
Section \ref{resultsSection}.  Models for the DM distribution and
scenarios for the enhancement of the gamma-ray flux are presented in
Section \ref{interpretationSection}.  We conclude in Section
\ref{susyLimitsSection} by discussing the implications of these
observations for the parameter space of allowed SUSY models.

\section{Review of Observational Targets}\label{sourceReviewSection}

Because the gamma-ray signature of the neutralino is proportional to
the square of the local density, the spatial scales that contribute
to the total gamma-ray flux from a DM halo are much smaller than the
spatial scales contributing to the halo mass.  On these small spatial
scales, the evolution of DM is typically driven by its interaction
with baryonic matter, which dominates the gravitational potential.
The influence of baryons in the form of dense stellar populations,
molecular clouds, and central black holes could potentially lead to a
much higher central DM concentration than that inferred from the
large-scale DM distribution and thus substantially enhance the
annihilation signal.  Therefore it is attractive to consider sources
that represent a diverse set of astrophysical environments, which
could boost the gamma-ray luminosity.  We have selected the dwarf
galaxies Draco and Ursa Minor, the local group galaxies M32 and M33,
and the globular cluster M15 for observations with the Whipple 10m
telescope based on the analysis of observational data and various
potential scenarios for DM enhancement in these objects.

\subsection{Draco and Ursa Minor}

Dwarf spheroidal galaxies have attracted considerable theoretical
attention as potential DM annihilation gamma-ray sources
\citep{baltz2000,tyler2002,strigari2007,bergstrom2006,colafrancesco2007}
due to their large observed mass-to-light ratios (M/L).  Studies of
surface brightness morphology have found a smooth symmetrical profile
in the core of Draco \citep{piatek2002,segall2007} but significant
structure in the central regions of Ursa Minor \citep{bellazzini2002},
which may be indicative of tidal interaction with the Milky Way.  The
observed stellar velocity dispersions in Draco and Ursa Minor imply
that the dynamics of these systems are dominated by DM on all spatial
scales and provide robust lower bounds on the astrophysical
contribution to the gamma-ray flux in these systems.

Both Draco and Ursa Minor possess low metallicity stellar populations
with an age of $\sim$10 Gyr \citep{aparicio2001,shetrone2001}.  The
absence of recent star formation suggests that neither system has
undergone a significant merger or accretion event since this early
star formation epoch.  Furthermore, because the two-body relaxation
time in these galaxies significantly exceeds the Hubble time, it is
unlikely that baryonic matter played a significant role in shaping the
present-day DM distribution.  Therefore primordial DM fluctuations
formed on small spatial scales during initial violent relaxation could
have been preserved and may boost the gamma-ray flux from these
objects.

\subsection{M15}

Although there is no observational evidence for the presence of
significant DM in globular clusters, the association of globular
clusters and DM halos fits naturally into the standard paradigm of
hierarchical structure formation.  In the primordial formation scenario
proposed by \citet{peebles1984}, globular clusters are formed in DM
overdensities in the early universe and may therefore retain a
significant fraction of this primordial halo in the current epoch.
Given that the extremely dense stellar cores of globular clusters
dominate the gravitational potential of these systems, the observable
effects of an extended DM halo may be minimal.  \citet{moore1996}
argued that the presence of tidal tails in some globular clusters
suggests that globular clusters are not embedded in DM halos.  Recent
simulations \citep{mashchenko2005a,mashchenko2005b,saitoh2006} have
challenged this picture, showing that an extended halo may be
compatible with the observable properties of globular clusters,
although much of the original halo mass could be stripped by tidal
interactions with the host galaxy.

The proximity and potentially high central DM density of M15 favor
this source for indirect DM searches.  With a core radius of $\sim$0.2
pc and extreme central density in excess of 10$^{7}$ M$_{\odot}$
pc$^{-3}$ \citep{dull1997}, M15 is the prototype for the
core-collapsed globular cluster.  During core collapse, the globular
cluster is predicted to relax through stellar two-body collisions to a
power-law density profile that extends down to the smallest observable
scales \citep{tremaine1987}.  If M15 was originally embedded in a DM
halo, this evolutionary process must significantly compress the
central DM distribution and dramatically enhance the gamma-ray flux.
However, the poorly understood process of kinetic heating of DM in the
core of the cluster by stars and hard binaries could lead to a
depletion of DM from this region.

\subsection{M32}\label{m32Section}

The model for compression of DM through the gravitational contraction
of dense stellar populations may also apply on the galactic scale.
M32 is the closest compact elliptical galaxy and may have formed in a
merging event between M31 and a low-luminosity spiral galaxy
\citep{bekki2001} in which the disk component of M32 was tidally
stripped.  Stellar kinematical data strongly support the presence of a
single supermassive compact object in the center of the galaxy with a
mass of 2--4 $\times$ 10$^6$ M$_\odot$ \citep{joseph2001}.  The core
of M32 has a relatively homogeneous stellar population with an
intermediate age of approximately 4 Gyr
\citep{corbin2001,delburgo2001}.  \citet{lauer1998} estimate M32's
core relaxation time scale to be 2--3 Gyr, implying that the nucleus
of M32 had at least a few relaxation times to evolve since the last
significant merging event.  Such events in which a massive black hole
binary is formed are predicted to deplete the central density by
evacuating stars and destroying any potential DM cusp in the galaxy
core \citep{milosavljevic2002}.  Collisional two-body relaxation of a
stellar population around a black hole is analytically predicted to
result in a steady state power-law stellar density profile with
power-law index between 3/2 and 7/4 \citep{bahcall1976}.  Optical and
infrared data indicate a stellar density profile compatible with a
power-law index in the range 1.4--1.9 at the resolution limit of 0.07
pc \citep{corbin2001,lauer1998}.  Because the condensation of baryons
in galactic nuclei may greatly enhance the central concentration of DM
halos, the stellar density in the core of M32, in excess of 10$^{7}$
M$_\odot$ pc$^{-3}$ and the highest known among nearby systems
\citep{lauer1998}, makes it a promising candidate for the detection of
DM annihilation.

\subsection{M33}\label{m33Section}

By observing astrophysical systems capable of rapid evolution, one may
be able to overcome dynamical limitations on the neutralino
annihilation rate, if it is limited by the scattering of WIMPs into a
very small annihilation region in the galactic nucleus.  M33 is
remarkable for the small relaxation time, $\sim$3 Myr, in its stellar
nucleus of approximately 0.2 pc, which results from the high stellar
density , 5$\times$10$^6$ M$_\odot$ pc$^{-3}$, and extremely low
velocity dispersion, 21 km s$^{-1}$, in this region \citep{lauer1998}.
M33 is a low-luminosity, DM dominated, bulgeless spiral galaxy with a
dark halo mass of approximately 5.1$\times$10$^{11}$ M$_\odot$
\citep{corbelli2000}. The mass of the black hole in its center is less
than 1.5$\times$10$^{3}$ M$_\odot$ \citep{gebhardt2001,merritt2001b}.
The stellar population in the nucleus of M33 can be modeled by two
bursts of star formation 2 and 0.5 Gyr ago suggesting the possibility
of a merger in the last $\sim$Gyr.  However, due to its rapid
collisional relaxation time, M33 could have developed a core-collapsed
nucleus in the period since the last merging event.

\section{DM Annihilation Flux}\label{dmFluxSection}

The differential flux of gamma rays from WIMP annihilation along a line
of sight is given by
\begin{equation}
  \frac{d\phi(\vec{\psi},\Delta\Omega)}{dE} = 
  \frac{\left<{\sigma}v\right>}{8{\pi}m_{\chi}^2}
  \left(\frac{dN(E,m_\chi)}{dE}\right)
  \int_{\Delta\Omega}d\Omega\int\rho^2\left(s,\vec{\psi},\Omega\right)ds,
\label{fluxeqn}
\end{equation}
where $\rho$ is the DM mass density, $m_{\chi}$ is the mass of the
WIMP particle, $\left<{\sigma}v\right>$ is the thermally averaged
product of the total self-annihilation cross section and the velocity
of the WIMP, $dN(E,m_\chi)/dE$ is the differential yield per
annihilation, $\Delta\Omega$ is the solid angle observed, and
$\vec{\psi}$ is the direction of the line of sight integration.

This expression can be factored as
\begin{equation}\label{diffFluxEqn}
\begin{array}{c}\displaystyle
  \frac{d\phi(\vec{\psi},\Delta\Omega)}{dE} = \phi_{1\%}
  \left(\frac{\left<{\sigma}v\right>}{3\times10^{-26} \
      \textrm{cm}^{3} \ \textrm{s}^{-1}}\right) \\[5mm]\displaystyle
  \times \left(\frac{100 \ \textrm{GeV}}{m_\chi}\right)^2
  \left(\frac{dN(E,m_\chi)/dE}{10^{-2} \ \textrm{GeV}^{-1}}\right)
  \frac{J(\vec{\psi},\Delta\Omega)}{1.45\times 10^{4}},
\end{array}
\end{equation}
where $\phi_{1\%}$ = 6.64 $\times$ 10$^{-12}$ cm$^{-2}$ s$^{-1}$ is
1\% of the integral Crab Nebula flux above 100 GeV as extrapolated
from the power-law fit of 3.2 $\times$ 10$^{-11}$ (E/TeV)$^{-2.49}$
cm$^{-2}$ s$^{-1}$ TeV$^{-1}$ reported by \citet{hillas1998}.
Following \citet{bergstrom1998b}, the astrophysical component, which
depends on the DM density profile, is expressed in terms of the
dimensionless quantity $J$,
\begin{equation}
  J(\vec{\psi},\Delta\Omega) = 
  \left(\frac{1}{\rho_{c}^2R_{H}}\right)
  \int_{\Delta\Omega}d\Omega\int\rho^2\left(s,\vec{\psi},\Omega\right)ds,
\label{jeqn}
\end{equation}
which we normalized to the critical density $\rho_{c} = 9.74 \times
10^{-30} \ \textrm{g} \ \textrm{cm}^{-3}$ and the Hubble radius $R_{H} =
4.16 \ \textrm{Gpc}$.

\subsection{DM Halo Profiles}\label{dmProfileSection}

Estimation of $J$ for a particular astrophysical object critically
depends on the DM density profile.  Numerical cold dark matter (CDM)
simulations, applicable in the regions where DM dominates the overall
gravitational potential, indicate the existence of a universal density
profile across the spectrum of halo masses, from dwarf galaxies to
galaxy clusters, which can be fit by,
\begin{equation}
\rho(r)=\rho_{s}\left(\frac{r}{r_s}\right)^{-\gamma}
\left(1+\left(\frac{r}{r_s}\right)^\alpha\right)^{-\frac{\beta-\gamma}{\alpha}}.
\label{cuspProfileEqn}
\end{equation}
Simulations have consistently found that the large-scale ($r \gg
r_{s}$) asymptotic behavior is compatible with $\beta$$\simeq$3.  The
value of the inner logarithmic slope, $\gamma$, is less certain due to
numerical resolution effects at the smallest scales.  The so-called
NFW profile, ($\alpha$,$\beta$,$\gamma$) = (1,3,1),
\citep{navarro1996} has an inner asymptotic of r$^{-1}$.  Profiles
with a steeper inner power-law cusp, ($\alpha$,$\beta$,$\gamma$) =
(1.5,3,1.5), have also been suggested \citep{moore1999b}.  Recent
high-resolution simulations have pointed to an intermediate value for
the inner power-law cusp $\gamma \sim$ 1.2 \citep{diemand2005b} and a
possible asymptotic shallowing approaching smaller scales
\citep{navarro2004,merritt2006}.

The density profiles of simulated DM halos can described by their
virial mass $m_{vir}$ and concentration $c$, which are related to
$\rho_{s}$ and $r_s$.  Using a sample of simulated DM halos with
masses 10$^{11}$--10$^{14}$ $h^{-1}$ M$_\odot$, \citet{bullock2001}
found that the median halo concentration at $z=0$ is correlated with
virial mass and can be well approximated by the expression,
\begin{equation}\label{concentrationEqn}
c = 9\left(\frac{m_{vir}}{1.5\times10^{13}h^{-1}M_{\odot}}\right)^{-0.13},
\end{equation}
with a scatter of $\Delta\log{c} \simeq 0.14$.   

Observations of low surface brightness galaxies \citep{deblok2001}
have motivated an alternative form for the halo density profile which
fits the rotation curves of these galaxies,
\begin{equation}\label{coredProfileEqn}
\rho(r)=\rho_{s}\left(1 + \frac{r}{r_s}\right)^{-\gamma}
\left(1+\left(\frac{r}{r_s}\right)^\alpha\right)^{-\frac{\beta-\gamma}{\alpha}}. 
\end{equation}
A specific choice of ($\alpha$,$\beta$,$\gamma$) = (2,3,1) is known
as the Burkert profile \citep{burkert1995}.  Given
the present state of uncertainty regarding the inner shape of DM
halos, we use both the NFW and Burkert profiles as representative of
the possible range of inner density asymptotes in DM dominated halos.

The existence of DM halo substructures is a central prediction
of CDM cosmology and may significantly enhance the DM annihilation
flux as compared to that predicted for a smooth halo.  Numerical
simulations have predicted that these substructures have a power-law mass
function dN/dM $\sim$ M$^{-\alpha}$ with $\alpha \simeq 2$
\citep{diemand2006}.  The mass spectrum cutoff, $m_{0}$, for these
structures is set by the free streaming and collisional damping length
scales in the early universe and depends on the microphysical
properties of the neutralino.  A neutralino with mass $m_\chi$ = 100
GeV is estimated to have $m_{0}$ $\sim$ 10$^{-6}$ M$_\odot$
\citep{green2005}.  The abundance of these substructures at $z=0$ is
uncertain and depends on the fraction that survive disruption during
the hierarchical merger and accretion processes.  Current CDM
simulations do not have sufficient resolution to model explicitly the
mass function at all relevant scales.  A recent estimate of the
contribution of substructure to the DM annihilation flux places a
lower limit on the enhancement to $J$ relative to a smooth halo of 2--3
\citep{diemand2007}.  \citet{strigari2007} obtain an upper bound of
$\sim$100 on the substructure enhancement by summing the contribution
of all substructures above $m_{0}$ $\simeq$ 10$^{-5}$ M$_\odot$ and
assuming a mass function with $\alpha$=1.9.

In the galaxies in which baryonic matter dominates the mass profile in
the central region, the observational constraints on the astrophysical
enhancement $J$ are weak.  Due to the large concentration of baryonic
matter on small scales, the evolution of DM in galactic nuclei is
likely significantly affected by the gravitational interaction of
neutralinos with a central black hole, high density stellar
populations, and gas.  Evolutionary condensation of baryons in the
core of a DM halo gives rise to a DM density enhancement beyond what
would be expected from the gravitational interaction of DM alone.
Using CDM simulations that combined both dissipationless DM particles
and a baryonic component composed of gas and stars, \citet{gnedin2004}
observed that baryon condensation increases the central concentration
of a DM halo.

The adiabatic compression of DM in the core of a halo through the slow
growth of a central black hole has been suggested as a mechanism for
increasing the flux from DM self-annihilation \citep{gondolo1999}.
However, the magnitude of this effect depends strongly on the ratio of
the initial and final masses of the black hole, its initial alignment
with the center of the DM halo, and the merging history of the
galactic nucleus \citep{ullio2001}.  \citet{merritt2001a} have shown
that a typical merger event between black holes of comparable mass
destroys cuspy profiles.  The effect of gravitational scattering of DM
particles by infalling baryons in the central regions of galactic
nuclei may also be significant.  The transfer of momentum to the dark
matter component in these interactions should lead to the partial
evaporation of DM from the center of the galaxy.  The ejection of DM
particles through the gravitational slingshot process in the vicinity
of binaries will further enhance the DM outflow from the centers of
galactic nuclei.  It might be expected that for a variety of
astrophysical objects the main contribution to $J$ originates from the
regions of high baryonic density.  The diversity of conditions
affecting the DM distribution in the core of a galactic nucleus
(presence or absence of a black hole, stellar velocity dispersion,
density of baryonic matter, merging history, etc.) argues for
observations of a variety of astrophysical objects in which DM
self-annihilation might be amplified due to the gravitational
interaction with baryons.

\subsection{DM Annihilation Spectrum}\label{annihilationSpectrumSection}

The differential yield of gamma-ray photons per neutralino
self-annihilation is a sum over final-state contributions,
\begin{equation}\label{spectrumEqn}
\begin{array}{rcl}\displaystyle
\frac{dN(E,m_\chi)}{dE} &=&\displaystyle 2b_{\gamma\gamma}\delta\left(E-m_\chi\right) \\
&+&\displaystyle \sum_i b_{{\gamma}i}\delta\left(E-m_\chi + \frac{m_{i}^2}{4m_\chi} \right)
+ \sum_i b_i\frac{dN_i(E,m_\chi)}{dE},
\end{array}
\end{equation}
where $b_{X}$ indicates the branching fraction of neutralino
self-annihilation into a specific final-state channel $X$.  The first
two terms represent annihilations into mono-energetic photons through
either one or two photon channels.  The last term is a sum over all
channels which contribute to the continuum flux, which arises
primarily from the decay of $\pi^0$ mesons produced in the
hadronization of the fermion and boson final states (for a review see
\citet{jungman1996}).  The two-body annihilation modes into
mono-energetic photons, $\gamma\gamma$ and $Z\gamma$, are
significantly suppressed as compared to the continuum component, with
branching ratios $\sim$10$^{-2}$--10$^{-3}$
\citep{bergstrom1997,bergstrom1998b}.

\begin{figure}[t]
\begin{center}
  \includegraphics[width=0.49\textwidth]{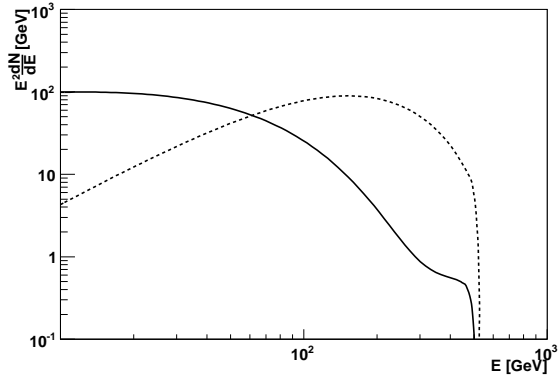}
  \caption{Spectral energy density per annihilation for a neutralino
    of mass 500 GeV annihilating to $b\bar{b}$ (solid line) and
    $\tau^{+}\tau^{-}$ (dashed line).  These spectra were generated
    using the \texttt{PYTHIA} Monte Carlo package
    \citep{sjostrand2001}.}
  \label{bbtautauspec}
\end{center}
\end{figure}

The differential spectrum of the $\pi^{0}$ decay component is
relatively featureless and similar for all channels.  It falls
exponentially at high energies terminating at $m_\chi$ where it is
enhanced by internal bremsstrahlung from charged decay components
\citep{birkedal2005}, resulting in an edge-like feature at E$_\gamma$
= m$_\chi$ (see Figure \ref{bbtautauspec}).  Decays into all quark and
bosonic states differ only slightly in the amplitudes of the $\pi^{0}$
and internal bremsstrahlung components.  However, decay into $\tau$
leptons generates a significantly harder spectrum, due to direct
production of $\pi^{0}$ mesons in processes such as $\tau^\pm
\rightarrow \pi^\pm\pi^0\nu$.  In this work, the $b\bar{b}$ and
$\tau^{+}\tau^{-}$ spectra are used to contrast the detection
prospects in the case of a soft or hard neutralino self-annihilation
spectrum, respectively.

Although observing regions of high baryonic density improves the
chances of indirect neutralino detection, the high density of baryonic
matter comes with the price of potentially high astrophysical
gamma-ray backgrounds.  High-energy processes that take place in
galactic nuclei, such as acceleration of particles in supernova shocks
or the jets formed by accreting black holes, interactions of cosmic
rays with molecular clouds, etc., can generate detectable gamma-ray
fluxes from these regions.  In these processes, particles producing
gamma rays are most likely accelerated stochastically resulting in a
power-law differential energy spectrum with index $\sim$2.  The
truncation of the spectrum of gamma rays from neutralino annihilation
at the neutralino mass can, in principle, be mimicked by traditional
astrophysical processes. For example, local absorption of gamma rays
through pair production in high-density optical and infrared diffuse
photon fields would cause an exponential cutoff in the spectrum.
However, such truncation would depend on the specific properties of
the source and would not be a common observational feature across
multiple objects.  

\section{Observations and Analysis}\label{dataSection}

\begin{deluxetable}{lcccc}
  \tablecaption{Summary of observation period and exposure time for ON
    and TRACK observing modes \label{dataObsTime} }

  \tablehead{
    &
    &
    \colhead{ON Exposure}&
    \colhead{TRACK Exposure}&
    \colhead{Total Exposure}\\
    \colhead{Source}&
    \colhead{Period}&
    \colhead{(hours)}&
    \colhead{(hours)}&
    \colhead{(hours)}
  }
  \startdata
  Draco&Mar 2003 - Jul 2003&7.4&6.9&14.3\\
  Ursa Minor&Jan 2003 - Jul 2003&7.9&9.3&17.2\\
  M32&Sep 2004 - Dec 2004&6.9&0&6.9\\
  M33&Oct 2002 - Dec 2004&7.9&9.2&17.0\\
  M15&Jun 2002 - Jul 2002&0.2&1.0&1.2\\
  \enddata
\end{deluxetable}

The TeV gamma-ray observations of the five sources reported in this
work were taken with the Whipple 10m telescope located at the Fred
Lawrence Whipple Observatory in southern Arizona, USA.  The telescope
is equipped with a camera consisting of 379 photomultiplier tubes,
covering a field of view of diameter 2.4$^\circ$, which detects the
short-duration Cherenkov light flashes emitted by secondary particles
generated in cosmic- and gamma-ray-induced atmospheric cascades.  A
detailed description of the telescope optics and camera configuration
is presented elsewhere \citep{kildea2007}.

\subsection{Observations}

The five sources were observed over the course of four observing
seasons using two different modes of operation: ON/OFF and TRACK.  The
ON/OFF mode is characterized by a sequential pair of 28 minute runs in
which the ON run is obtained by pointing the telescope in the
direction of the source and the OFF run is taken at the same azimuth
and elevation but offset by 28 minutes in right ascension for
background estimation.  The ON/OFF technique minimizes systematic
errors due to the changing state of the atmosphere and variations in
the night-sky brightness.  The TRACK mode observations are not
followed by dedicated OFF observations.  Instead, a contemporaneous
but unrelated OFF run of similar elevation and average night-sky
background noise is selected and then analyzed with the TRACK mode
observation in the same way as the ON/OFF pair.  To reduce systematic
errors due to variations in the night-sky background between ON and
OFF fields, artificial noise was injected on a pixel by pixel basis
into either ON or OFF runs to equalize them \citep{weekes1996}.  The
total accumulated exposures in ON and TRACK modes and observation
epochs for each source are summarized in Table \ref{dataObsTime}.

\subsection{Data Analysis}\label{dataAnalysisSection}

The light distributions of the Cherenkov images were parameterized
using a standard moment analysis \citep{hillas1985}.  The parameter
\textit{size} characterizes the total amount of light in the image
while the parameters \textit{width} and \textit{length},
\textit{distance}, and $\alpha$ describe the shape, location, and
orientation of the image, respectively.  A set of selection criteria,
referred to as \textit{Supercuts} \citep{reynolds1993}, were used to
select gamma-ray-like images with an efficiency of $\sim$50\%, while
rejecting $>$99\% of cosmic rays.

For events passing the \textit{Supercuts} criteria, a histogram of the
$\alpha$ parameter is generated.  Gamma-ray events originating from a
source in the center of the field appear as an excess in the region of
the $\alpha$-histogram with $\alpha \leq$ 15$\degree$.  The residual
background in the signal region of the ON-source observation is
estimated from the number of events obtained from the same region of
the OFF-source observation rescaled by the ratio of the OFF to ON
livetimes.  In the case of the TRACK mode observations, the background
estimate from the matching OFF-source observation is scaled by an
additional factor, equalizing backgrounds in a sideband region
20\degree \ $< \alpha <$ 65\degree \ on a run-by-run basis.  For each
source, the significance of an excess in the signal region is
evaluated with the maximum likelihood method of \citet{li1983}.

\begin{figure}[t]
  \begin{center}

    \includegraphics[width=0.49\textwidth]{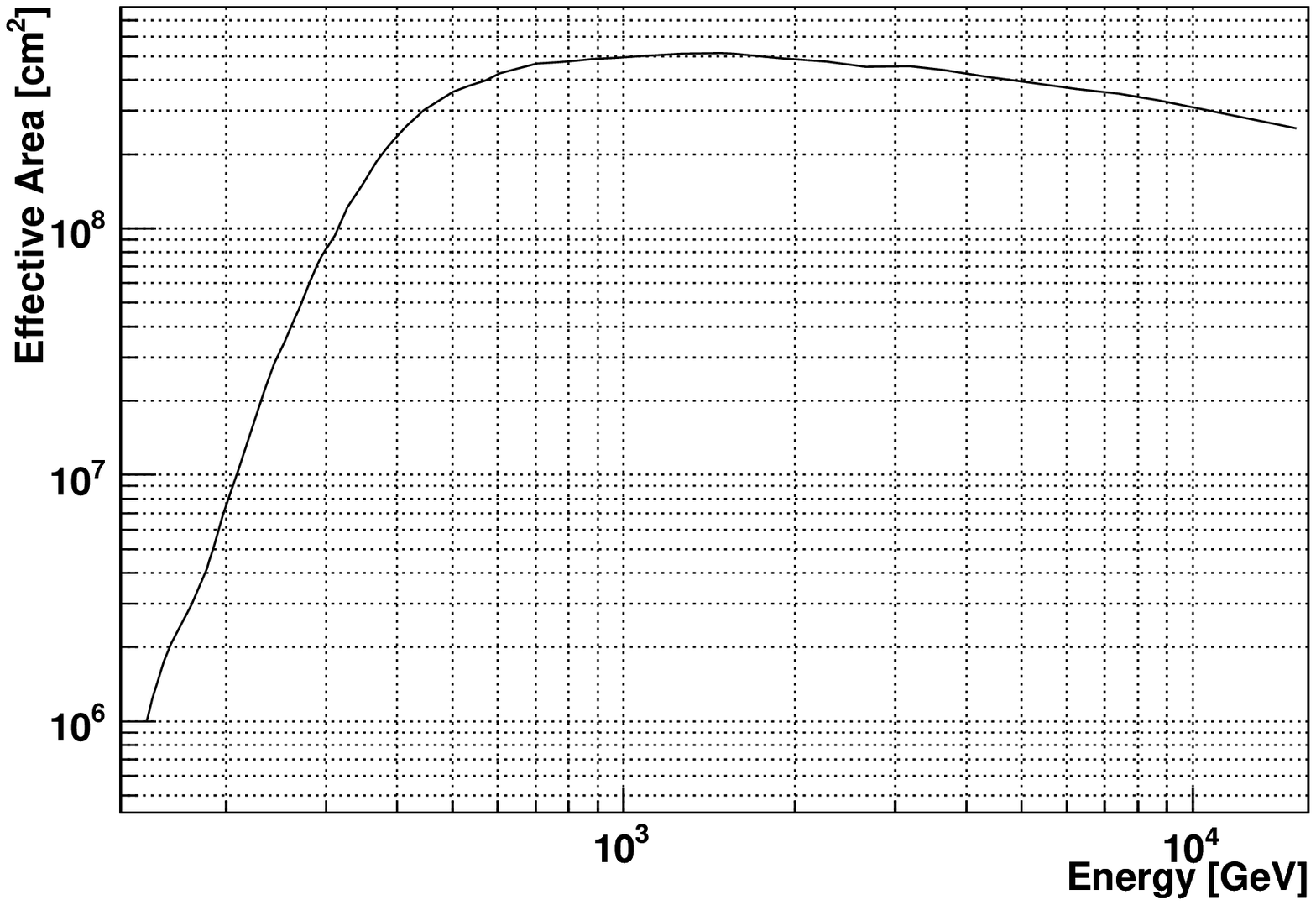}
    \caption{Effective area of the Whipple 10m telescope as a function
      of energy after selection of gamma-ray-like events with the
      \textit{Supercuts} criteria.  The effective area below 150 GeV
      was assumed to be 0 in all computations. }
    \label{whippleEffArea}
    \includegraphics[width=0.49\textwidth]{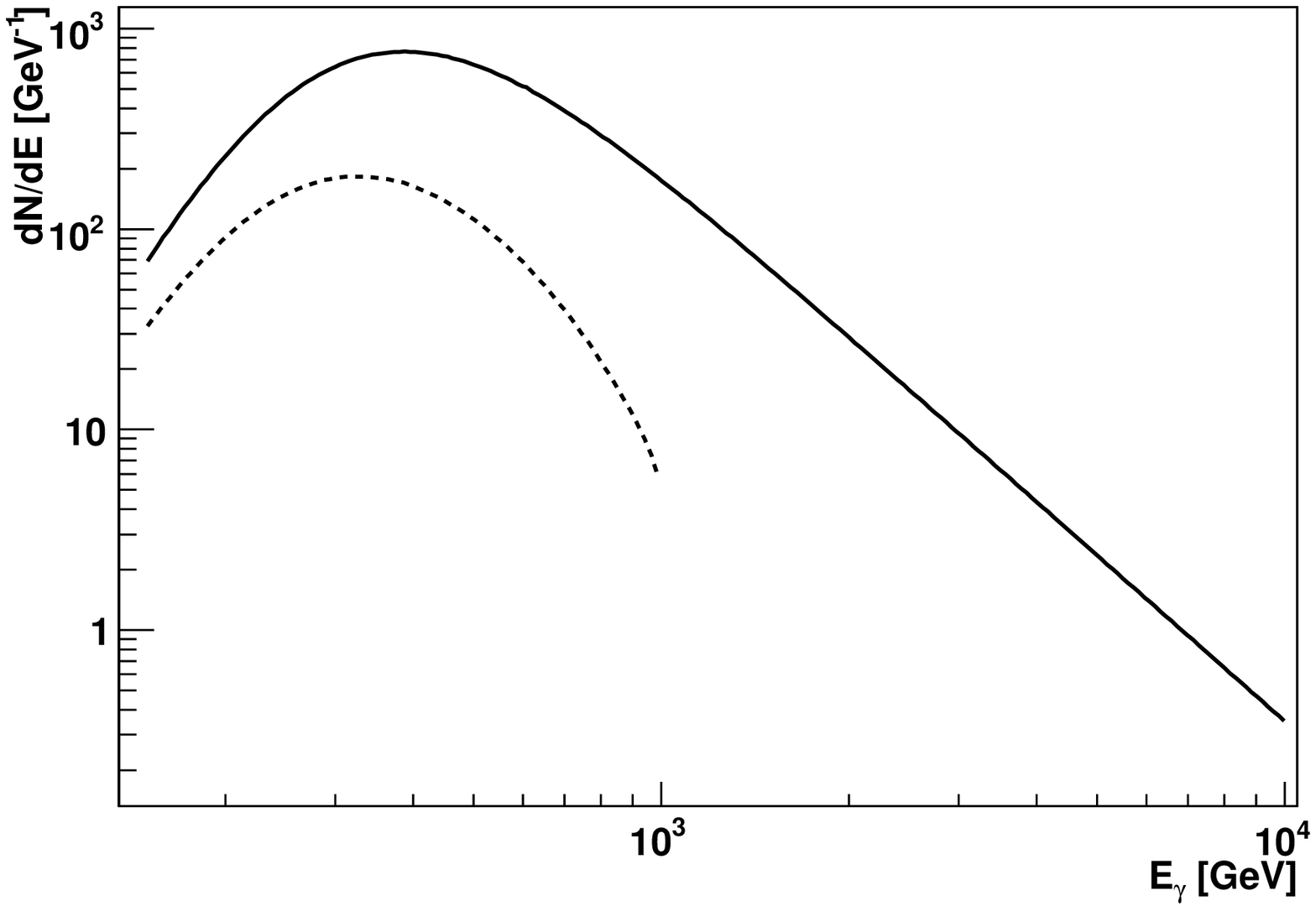}
    \caption{Comparison of the differential detection rate of the Crab
      Nebula (solid line) and a DM halo with
      $\left<{\sigma}v\right>=3\times10^{-26}$, $m_\chi=$ 1 TeV, and
      $J=10^{6}$ (dashed line).  The \textit{Supercuts} selection
      criteria are used.}
    \label{crabDiffRate}

  \end{center}
\end{figure}

The method of analysis described estimates the integral gamma-ray
excess which is the convolution of the effective area of the Whipple
10m telescope shown in Figure \ref{whippleEffArea} with the spectrum
of the source.  The gamma-ray collecting area is nearly constant at
$\sim$4 $\times 10^{8}$ cm$^{2}$ for photon energies above 1 TeV and
declines rapidly in the energy regime below $\sim$400 GeV.  The
differential gamma-ray detection rate, the product of effective
collecting area and source spectrum, is shown for a Crab Nebula-like
spectrum in Figure \ref{crabDiffRate}.  The peak of the detection rate
is $\sim$400 GeV and is a relatively weak function of the index of the
power-law spectrum in the range $\sim$2--3.  Despite the fact that the
continuum neutralino spectrum can not be approximated as a power law,
for neutralino masses exceeding approximately 400 GeV the differential
gamma-ray detection rate peaks in the vicinity of this energy as
illustrated in Figure \ref{crabDiffRate} for $m_\chi =$ 1 TeV.  Thus
the integral constraints discussed in this paper are appropriate for a
source with a neutralino-like spectrum as they have been optimized to
produce the best sensitivity for the sources which peak in the energy
range 300--400 GeV.  The integral method remains applicable for
neutralino masses below 400 GeV. However, the sensitivity deteriorates
rapidly as the peak of the detection rate falls significantly below
400 GeV.

In this analysis we have not searched for the monoenergetic line
feature of the neutralino self-annihilation spectrum.  The Whipple 10m
telescope has an energy resolution of $\sim$30\% which ultimately
limits the sensitivity to monoenergetic photons.  The search for the
continuum component of the spectrum can provide an equivalent
or better sensitivity to neutralino annihilations as compared to
searching for the monoenergetic line given that under the preferred
particle physics scenarios the branching ratios to two and
one photon channels is less than one percent. A dedicated search for
monoenergetic photons could be a goal of a refined analysis if an
integral excess is detected.  

The accuracy of the reconstruction of the arrival direction of
individual photons for the Whipple 10m telescope for a source in the
center of the field of view is a function of photon energy and changes
from $\sim$20$^\prime$ at 300 GeV to $\sim$2$^\prime$ at 10 TeV.  The
event selection criterion $\alpha\leq$15$\degree$ corresponds to an
angular cut in the image plane of $\sim$15$^\prime$ which translates
into a solid angle $\Delta\Omega$ of 6 $\times$ 10$^{-5}$ sr which is
used in the calculation of $J$ (Equation \ref{jeqn}).

The Crab Nebula is the standard calibration source for ground-based
gamma-ray astronomy.  Observations of the Crab Nebula were taken
during the same four year period as the observations of the sources of
interest.  These data were analyzed with the same technique to
calibrate the gamma-ray detection efficiency of the Whipple 10m
telescope.  The differential flux of the Crab nebula at 400 GeV is
estimated using the spectral parameterization of \citet{hillas1998}.
By scaling the observed gamma-ray rate from the putative DM sources to
the contemporaneous rate of the Crab Nebula, the systematics due to
the changing performance of the telescope optics and camera system
were corrected for.

\subsection{Results}\label{resultsSection}
\begin{deluxetable}{lccccc}
  \tablecaption{Detected Gamma-ray rates, inferred upper limits on the
    gamma-ray rate, significances, and differential gamma-ray flux upper
    limits \label{rateSigTable}}

  \tablehead{
    &
    \colhead{Excess}&
    \colhead{95\% C.L. Upper Limit}&
    &
    \multicolumn{2}{c}{$E^{2}\frac{dF}{dE}$ at 400 GeV Upper Limit}\\
    \colhead{Source}&
    \colhead{($\gamma$ min$^{-1}$)}&
    \colhead{($\gamma$ min$^{-1}$)}&
    \colhead{$\sigma$}&
    \colhead{(erg cm$^{-2}$ s$^{-1}$)}&
    \colhead{(\% Crab)}
  }

  \startdata
  Draco&\phm{-}0.001 $\pm$ 0.066&0.14&\phm{-}0.02&
  5.10 $\times$ 10$^{-12}$&6.23\\
  Ursa Minor&\phm{-}0.075 $\pm$ 0.070&0.20&\phm{-}1.07&
  7.30 $\times$ 10$^{-12}$&8.94\\
  M32&-0.240 $\pm$ 0.170&0.21&-1.44&
  6.00 $\times$ 10$^{-12}$&7.34\\
  M33&-0.012 $\pm$ 0.085&0.16&-0.14&
  5.75 $\times$ 10$^{-12}$&7.04\\
  M15&\phm{-}0.496 $\pm$  0.269&0.94&\phm{-}1.80&
  2.68 $\times$ 10$^{-11}$&32.8\\
  \enddata
\end{deluxetable}

Using the \textit{Supercuts} criteria, the excess gamma-ray rate and
its significance for each source was calculated under the hypothesis
of a point source in the center of the field of view.  No significant
excess was detected from any of the five sources observed (see Table
\ref{rateSigTable}).  In the absence of a significant detection, the
95\% C.L. upper limit on the rate from each source was calculated
following the method of \citet{helene1983}.  The upper limit on the
differential spectral energy density $E^{2}dF/dE$ at 400 GeV is
calculated for each source by scaling the gamma-ray rate to the
observed Crab rate during the same observation epoch.

The 95\% C.L. upper limits for all sources are 6--9\% of the Crab
Nebula flux with the exception of M15 for which the exposure time was
significantly shorter.  ACT observations of M32 were reported by the
HEGRA collaboration \citep{aharonian2003a}.  This work places a 99\%
C.L integral flux upper limit of 4.4\% of the Crab Nebula rate for M32
which is compatible with the results presented in this work.

\section{Analysis of Astrophysical Enhancement Factors}
\label{interpretationSection}

The primary difficulty in constraining the parameter space of
allowable SUSY models is due to the significant uncertainty in the
astrophysical enhancement factor, $J$ (Equation \ref{jeqn}).  A lower
bound on $J$ can be estimated by extrapolating the DM density measured
on large spatial scales ($r \gg r_{s}$) into the small scales (where
most of the DM annihilation signal originates) by using a profile with
a weak cusp (NFW) or central core (Burkert).  However, for each of the
sources considered there may exist astrophysical mechanisms which
could enhance the density of DM in the core of the halo and boost the
luminosity due to neutralino self-annihilation by several orders of
magnitude.  A discussion follows for each source which presents both a
conservative estimate of $J$ as well as possible scenarios for its
enhancement which take into account the source's unique astrophysical
environment.

\subsection{Draco and Ursa Minor}

\begin{figure}[t]
  \begin{center}
    \includegraphics[width=0.49\textwidth]{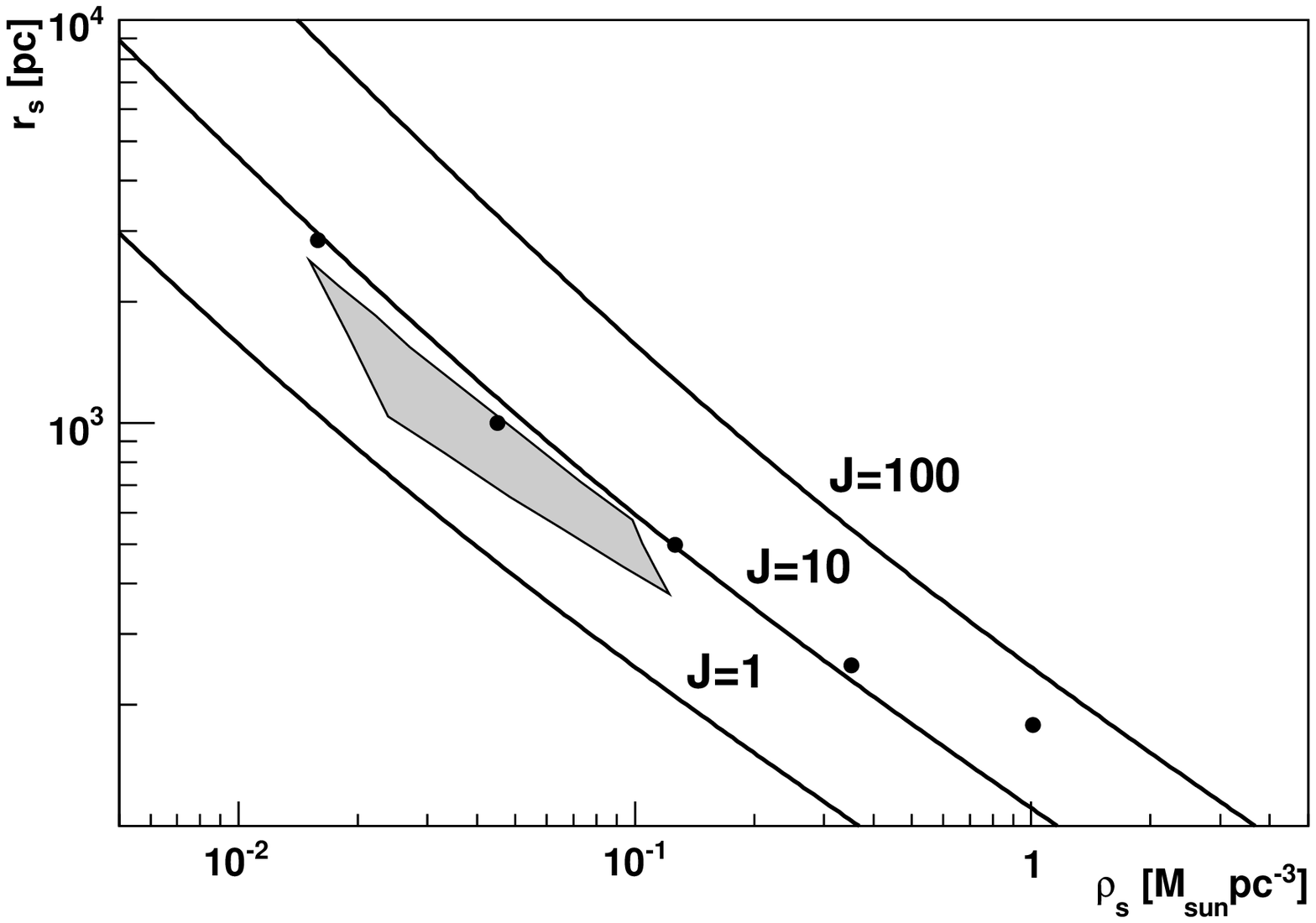}
    \caption{Constraints from stellar kinematics data on the
      parameters $r_{s}$ and $\rho_{s}$ of the Draco DM halo under the
      assumption of an NFW profile.  The gray contour and solid
      circles indicate the best-fit mass models of \citet{strigari2007} and
      \citet{mashchenko2006}, respectively.  The parameters $r_{s}$ and
      $\rho_{s}$ are defined for the NFW and Burkert halo profiles in
      Equations \ref{cuspProfileEqn} and \ref{coredProfileEqn}.}
    \label{dracoNFWProfile}
  \end{center}
\end{figure}

\begin{figure}[t]
  \begin{center}
    \includegraphics[width=0.49\textwidth]{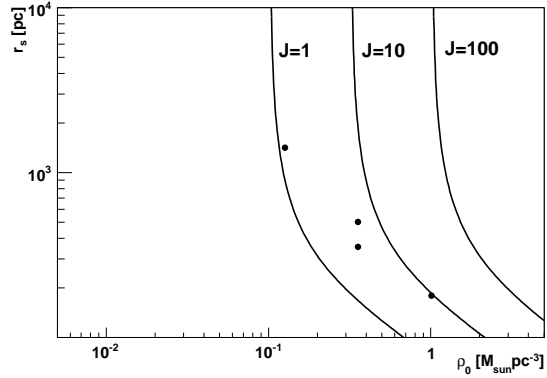}
    \caption{Constraints from stellar kinematics data on the
      parameters $r_{s}$ and $\rho_{s}$ of the Draco DM halo under the
      assumption of a Burkert profile.  The solid circles indicate the
      best-fit mass models of \citet{mashchenko2006}.  Thick solid lines
      indicate contours of constant $J$.}
    \label{dracoBurkertProfile}
  \end{center}
\end{figure}

\begin{figure}[t]
  \begin{center}
    \includegraphics[width=0.49\textwidth]{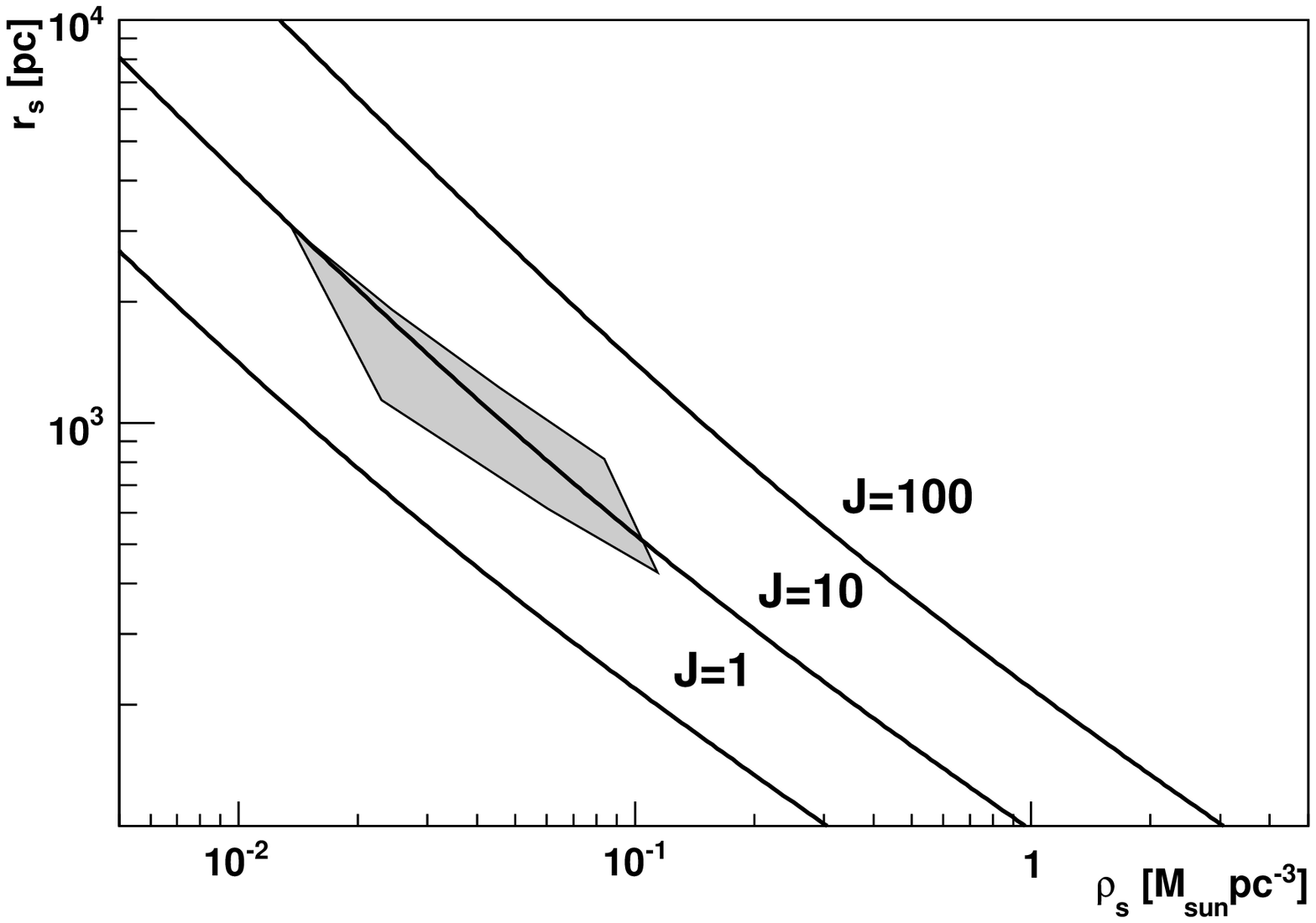}
    \caption{Constraints from stellar kinematics data on the
      parameters $r_{s}$ and $\rho_{s}$ of the Ursa Minor DM halo
      under the assumption of an NFW profile.  The grey contour
      indicates the best-fit mass models of \citet{strigari2007}.  Thick
      solid lines indicate contours of constant $J$.  }
    \label{ursaNFWProfile}
  \end{center}
\end{figure}

Because the gravitational potentials of Draco and Ursa Minor are
dominated by DM on all observationally resolved scales, studies of
stellar kinematics in these systems provide robust constraints on
their DM density profiles at radii greater than $\sim$0.5 kpc.
\citet{strigari2007} have used the radial velocity data sets compiled
by \citet{wilkinson2004} and \citet{munoz2005} to derive constraints
on the parameters $r_{s}$ and $\rho_{s}$ of their DM halos under the
assumption that the DM follows an NFW density profile.  The best-fit
contours in the $r_{s}$--$\rho_{s}$ plane are shown in Figures
\ref{dracoNFWProfile} and \ref{ursaNFWProfile} for Draco and Ursa
Minor, respectively.  A similar analysis of the Draco data set by
\citet{mashchenko2006} which considered both NFW and Burkert DM
density profiles obtained similar constraints on $r_{s}$ and
$\rho_{s}$, shown in Figures \ref{dracoNFWProfile} and
\ref{dracoBurkertProfile}.  Both analyses find a region of degeneracy
in the $r_{s}$--$\rho_{s}$ plane which is attributable to the weak
constraints on the stellar velocity anisotropy.  The region of
degeneracy for these models, however, is nearly parallel to the
isocontours of $J$ which results in a much smaller uncertainty for
this parameter.  Conservative allowable ranges for $J$ calculated using
the mass models presented by \citet{mashchenko2006} and
\citet{strigari2007} are summarized in Table \ref{jtable}.

The inner logarithmic slope ($\gamma$) of the DM density profile in
Draco and Ursa Minor is not observationally constrained.  A value of
$\gamma > 1$ could potentially enhance by 2--3 orders of magnitude the
value of $J$ with respect to the estimates presented in Table
\ref{jtable}.  The presence of DM substructures is another potential
factor contributing to the enhancement of $J$.  \citet{strigari2007}
calculated an upper bound of $\sim$100 on the enhancement due to
substructures for a generic DM halo.  The distortion of the
gravitational potential by DM substructures may be reflected in the
distribution of stellar populations.  Stellar lumps within the central
10 arc minutes of Ursa Minor have been detected, and the study of the
stellar proper motion suggests that the lifetime of these structures
should be no longer than $\sim$5 Myr \citep{eskridge2001}.  If proven
statistically significant, these observations may indicate the
existence of small scale substructures in the DM distribution.
However a conventional astrophysical explanation such as the
projection of a cold extratidal population of stars is also possible
\citep{wilkinson2004}.

\subsection{M15}

Models for the M/L of M15 are consistent with a purely baryonic mass
profile \citep{vandenbosch2006} and therefore indicate that the mass
of a putative DM halo must be significantly less than the mass of the
stellar component within the observable extent of the cluster.
However, given the compact nature and extreme central density of M15,
this constraint is not a significant limitation on the potential DM
annihilation signal, as evident from the following estimates.

Consider a hypothetical halo model for M15 parameterized by a virial
mass $m_{vir}$ and concentration parameter $c$.  The virial mass is
unlikely to be significantly less than estimates of the present
baryonic mass of the cluster of $\sim$5$\times$10$^{5}$ M$_\odot$
\citep{dull1997,mcnamara2004}.  Conversely, because of its distance
from the Milky Way, the halo mass can not be significantly greater
than $\sim$10$^{8}$ M$_\odot$, or dynamical friction would have
resulted in the inspiral of the cluster in less than a Hubble time.
Based on these constraints, we adopt a range for $m_{vir}$ of
5$\times$10$^{6}$ M$_\odot$--5$\times$10$^{7}$ M$_\odot$ and
consequently a range for $c$ of 32--82 as estimated with the
\citet{bullock2001} relationship, Equation \ref{concentrationEqn}.
Although the $c(m_{vir})$ correlation was derived with simulated halos
of mass $\sim$10$^{11}$--10$^{14}$, \citet{colin2004} have found that
this relationship holds for simulated halos with masses down to
$\sim$10$^{7}$ M$_\odot$.  To account for tidal disruption due to the
interaction with the Milky Way, we truncate the hypothetical model of
the DM profile at the optical extent of the cluster, $\sim$30 pc.

\begin{figure}[t]
  \begin{center}
    \includegraphics[width=0.49\textwidth]{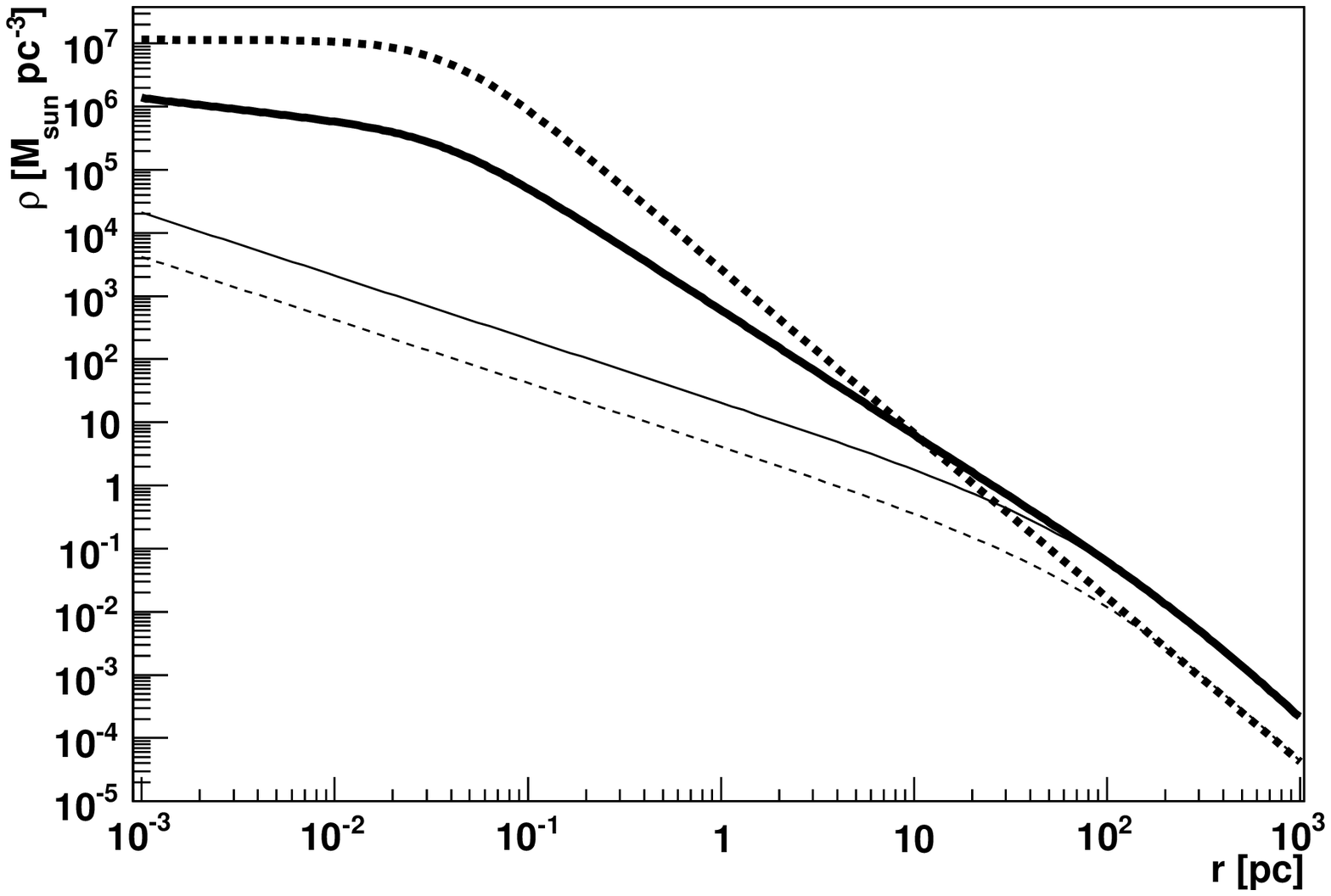}
    
    \caption{ Shown is the M15 DM density profile before (thin solid line)
      and after (thick solid line) adiabatic compression modeled as
      described in the text.  Thin and thick dashed lines show the same
      comparison for the baryonic density profile.  }
    
    \label{m15Profile}
  \end{center}
\end{figure}

\begin{figure}[t]
  \begin{center}
    \includegraphics[width=0.49\textwidth]{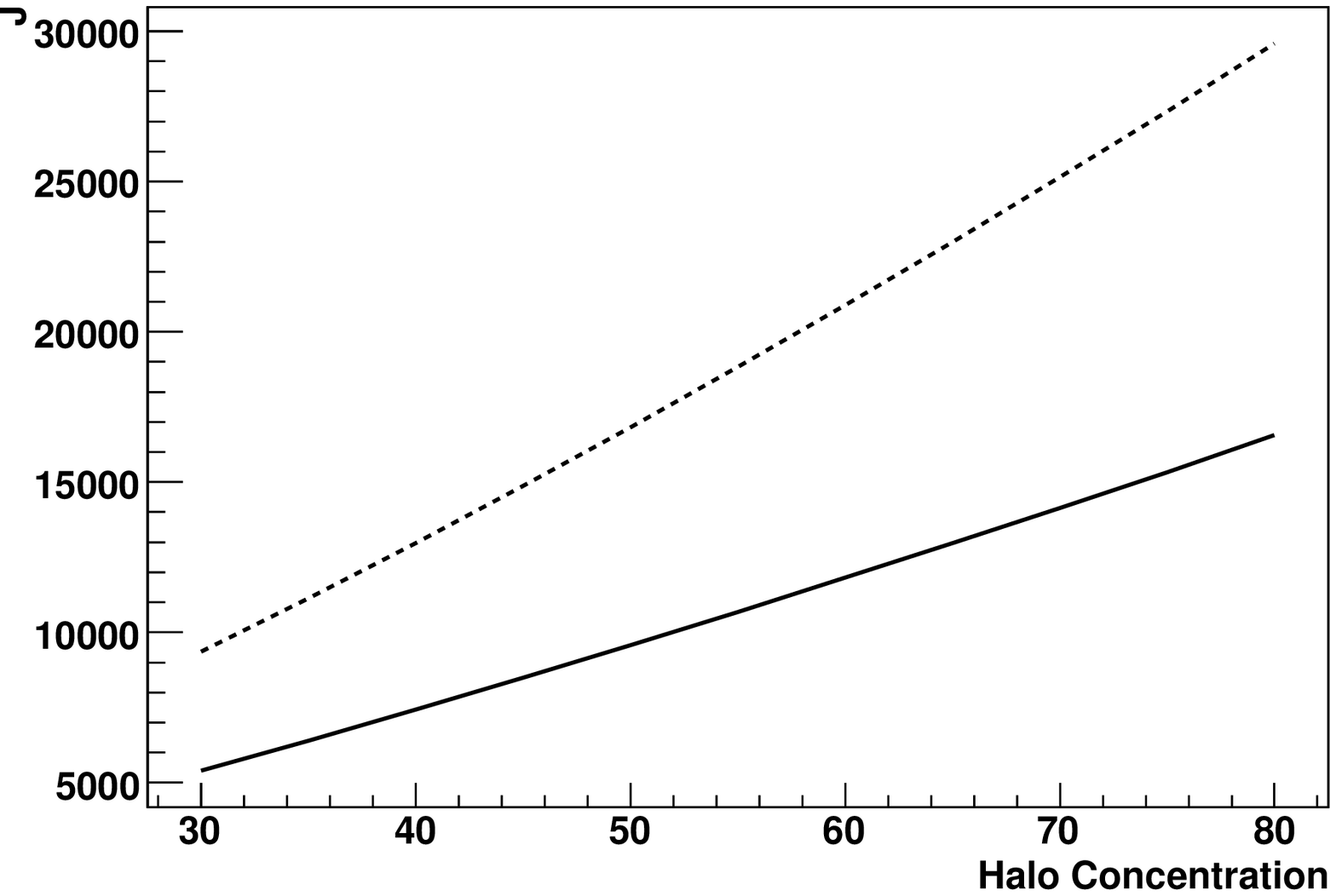}

    \caption{$J$ for the NFW+AC model of M15 as a function of the
      concentration parameter of the initial DM halo.  The solid and
      dashed lines show the scaling for a DM halo of total virial mass
      5$\times$10$^{6}$ M$_\odot$ and 5$\times$10$^{7}$ M$_\odot$
      respectively. }
    
    \label{m15JPlot}
  \end{center}
\end{figure}

For the adopted ranges of $m_{vir}$ and $c$, we estimate $J$ to be
7--150 if the DM halo follows an NFW profile.  However, the adiabatic
compression of DM in the core of M15 could further enhance the
annihilation signal by several orders of magnitude.  Simulations by
\citet{mashchenko2005b} of a two-component globular cluster with stars
and DM have demonstrated an enhancement to the central DM density as
the baryonic mass profile becomes more centrally concentrated through
two-body interactions.  To evaluate the enhancement to the central DM
density for our hypothetical model of M15, we have used the adiabatic
compression model of \citet{blumenthal1986}.  With the assumption that
the DM travels on circular orbits, this model relates the initial and
final baryon and DM mass profiles by the equation,
\begin{equation}
\begin{array}{c}\displaystyle
  \left[M_{DM,i}(r_{i}) + M_{b,i}(r_{i})\right]r_{i} = \\\displaystyle
  \left[M_{DM,f}(r_{f}) + M_{b,f}(r_{f})\right]r_{f}.
\end{array}
\end{equation}  
The initial distribution of DM is described by an NFW profile
parameterized by $m_{vir}$ and $c$, while the initial baryonic mass
profile is assumed to follow the DM mass profile with the cosmological
baryon-to-DM ratio of 0.2, following the assumption that globular
clusters are among the oldest gravitationally bound systems with cores
that have not been significantly influenced by merger events during
their evolution.  For the final baryonic mass distribution, we adopt a
cored profile with ($\alpha$,$\beta$,$\gamma$) = (2,2.6,0), $r_{c}$ =
0.04 pc, and $\rho_{s}\sim$10$^{7}$ M$_\odot$ pc$^{-3}$ which
approximates the nonparametric stellar mass profile presented in
\citet{gebhardt1997}.  The predicted profile after adiabatic
compression, hereafter denoted as NFW+AC, is shown in Figure
\ref{m15Profile} for the case of $m_{vir}$ = 10$^{7}$ M$_{\odot}$ and
$c$ = 50.  The density of DM interior to $\sim$10 pc is enhanced by a
factor $\sim$10--10$^{2}$ resulting in an increase in $J$ of
$\gtrsim$10$^{2}$.  The truncation of the DM halo at 30 pc has a
negligible effect on the total annihilation signal.  Figure
\ref{m15JPlot} illustrates the scaling of $J$ with the assumed virial
mass and concentration of the DM halo.  The range of $J$ for the
models with and without adiabatic compression are summarized in Table
\ref{jtable}.

\subsection{M32}

The stellar surface brightness profile of M32 is characterized by a
bulge with a half-light radius of 100 pc and a faint surface
brightness excess beyond 300 pc which could be interpreted as the
remnant of a tidally stripped disk \citep{graham2002}.  A V-band M/L of
2.51 interior to $\sim$50 pc \citep{vandermarel1998} is consistent
with the expectations of an intermediate age stellar population.
Due to the absence of a significant disk component, the available
kinematical data do not constrain the presence of an extended DM halo
in this system.  Its proximity to M31 has lead to speculation that M32
may be the remnant of a normal elliptical or late-type spiral galaxy
which was tidally stripped as it passed through the disk of M31
\citep{bekki2001,choi2002}.  The extent to which tidal interactions
may have stripped the DM halo in this event is unclear.

\begin{figure*}[t]
  \begin{center}
    \includegraphics[width=0.49\textwidth]{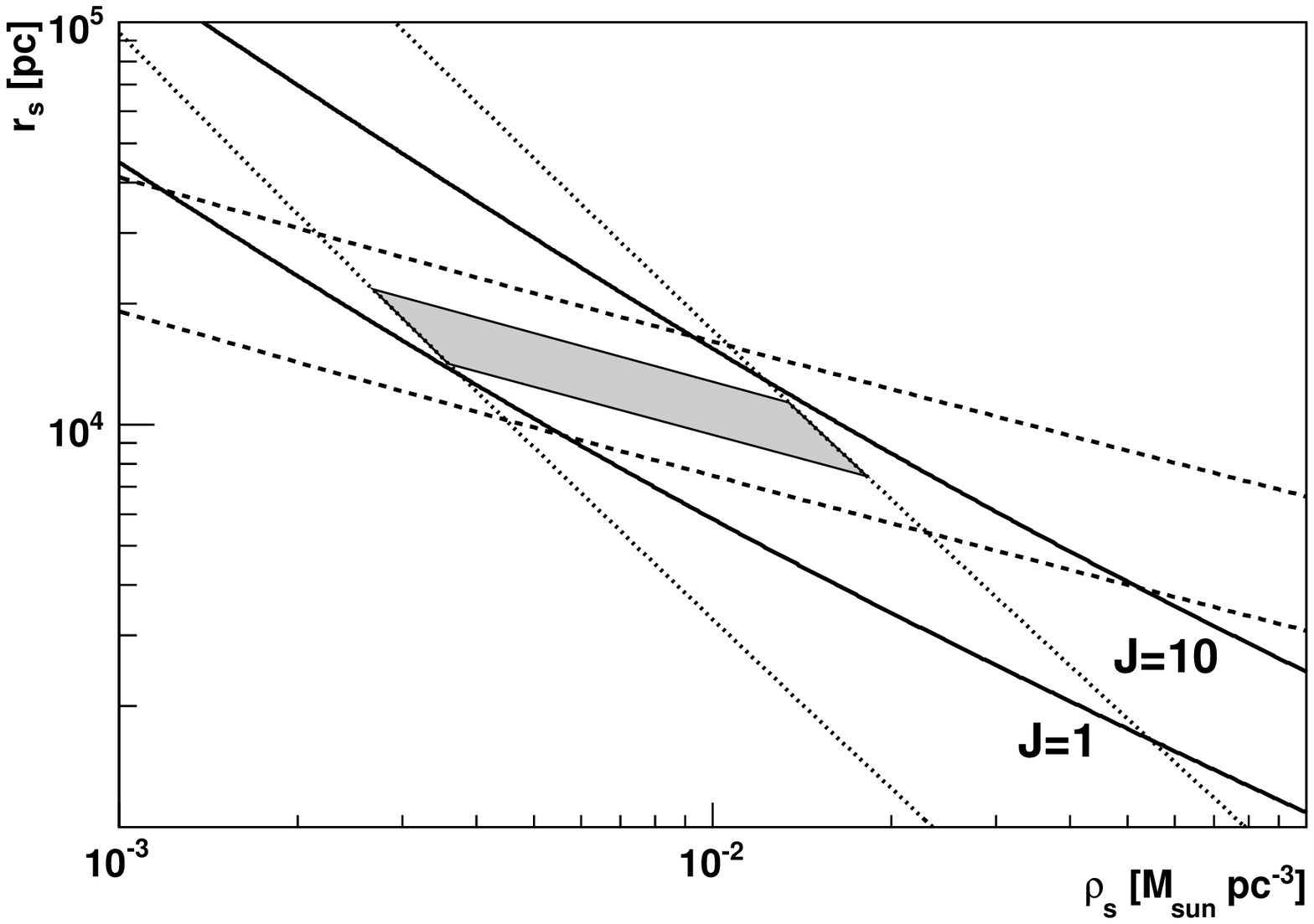}
    \includegraphics[width=0.49\textwidth]{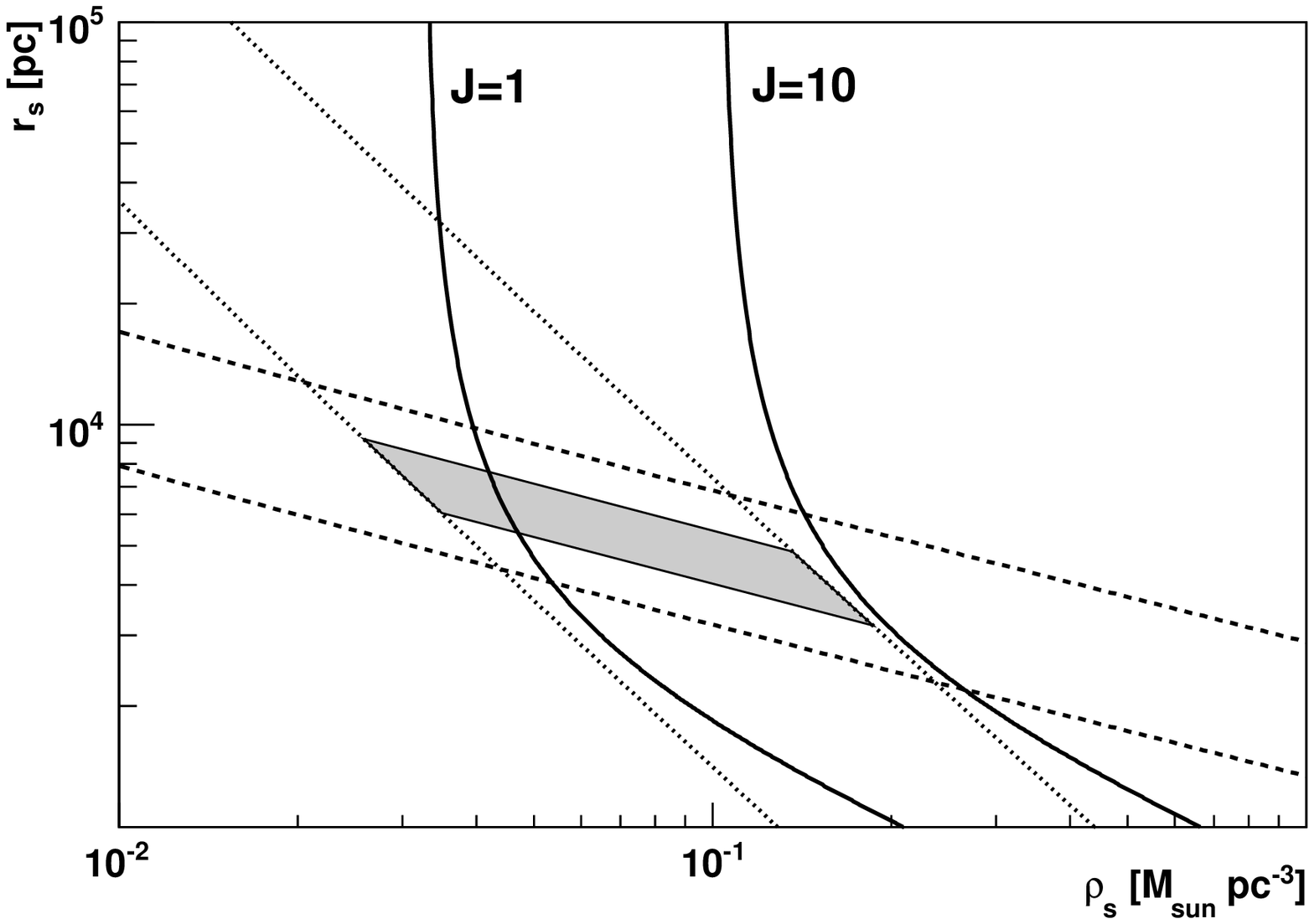}
    
    \caption{Constraints derived using Equations \ref{virialMassEqn}
      and \ref{concentrationEqn} for M32 on the parameters $r_{s}$ and
      $\rho_{s}$ of its DM halo under the assumption of an NFW (left
      panel) and Burkert (right panel) profile.  Dashed lines show
      contours of constant virial mass for 10$^{11}$ M$_\odot$ (lower)
      and 10$^{12}$ M$_\odot$ (upper). Dotted lines denote the $\pm$1
      standard deviation bounds on the median halo concentration to
      virial mass relation, Equation \ref{concentrationEqn}. Thick
      solid lines indicate contours of constant $J$.}

    \label{m32JPlot}
  \end{center}
\end{figure*}

The mass of the M32 DM halo can be estimated under the hypothesis that
its progenitor was a late-type spiral by using the correlation between
bulge velocity dispersion $\sigma_{c}$ and maximum circular velocity
$v_{c}$ presented by \citet{ferrarese2002} for a sample of nearby
spiral galaxies.  \citet{ferrarese2002} notes that $v_{c}$ may be
related to the virial mass of the DM halo, $m_{vir}$, using the
correlation between them observed for simulated halos by
\citet{bullock2001}.  By combining these two correlations the
following relation between the bulge velocity dispersion and DM halo
mass is obtained:
\begin{equation}
\frac{m_{vir}}{10^{12} M_\odot}\sim
5.6\left(\frac{\sigma_{c}}{200 \ \textrm{km/s}}\right)^{2.79}.
\label{virialMassEqn}
\end{equation}
The average bulge velocity dispersion in M32 of 76 $\pm$ 10 km
s$^{-1}$ \citep{vandermarel1998} suggests a virial halo mass of
2.5--5$\times$10$^{11}$ M$_\odot$.  By further assuming that the DM
halo of M32 follows an NFW or Burkert profile, the scale radius and
normalization of the halo are fixed by a choice of the concentration
parameter $c$, predicted to be in the range 10--20 by Equation
\ref{concentrationEqn}.  Constraints on the parameters for both NFW
and Burkert DM density profiles are plotted in Figure \ref{m32JPlot}.
The estimated range of $J$ for these mass models is found to be 1--10
(see Table \ref{jtable}).  The models with the range of $m_{vir}$
and $c$ considered here are compatible with the the observed M/L in the
interior of M32.

\begin{figure}[t]
  \begin{center}
    \includegraphics[width=0.49\textwidth]{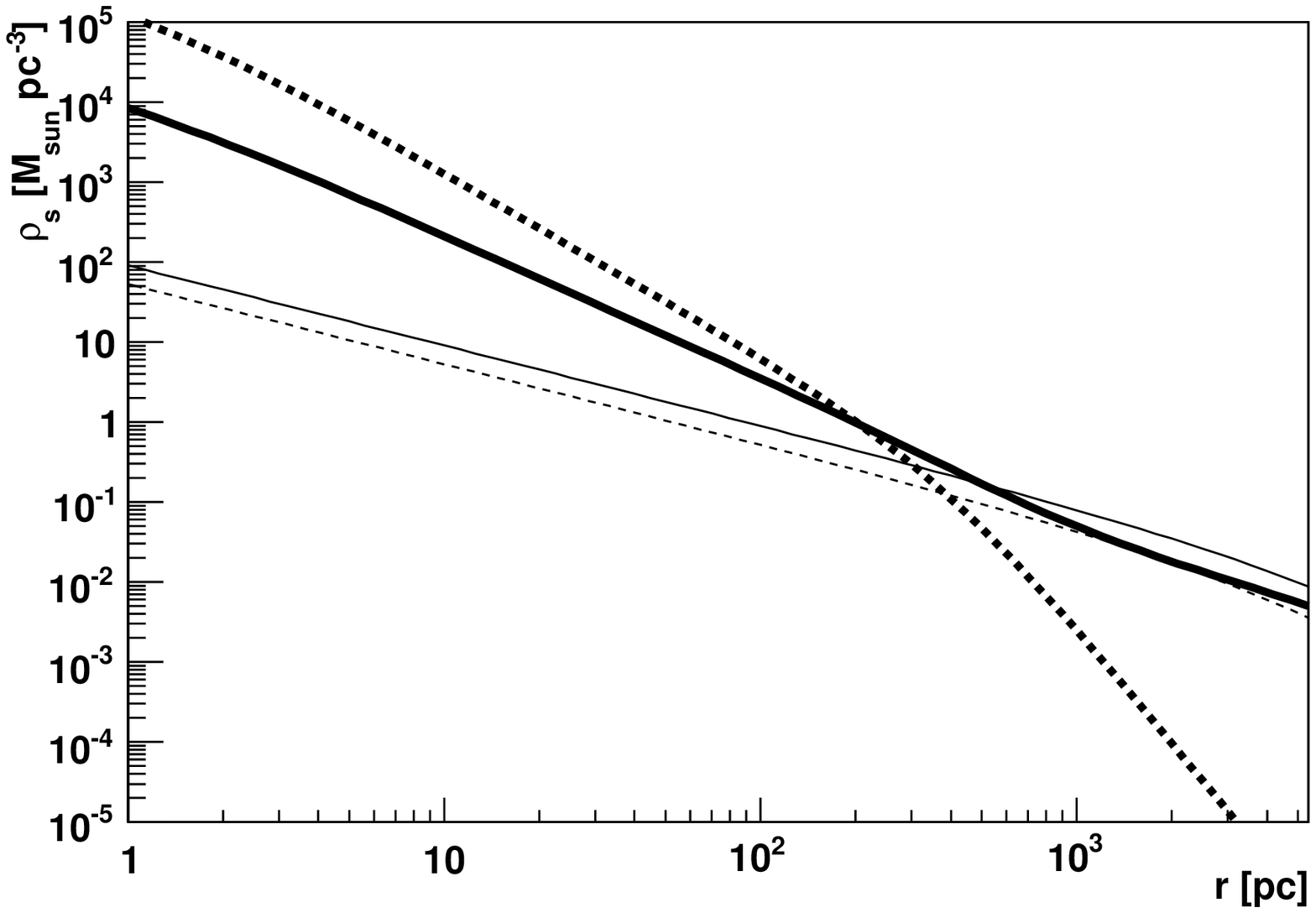}
    
    \caption{A comparison for M32 of the stellar density profile of
      \citet{vandermarel1998} (thick dashed line) with the modeled DM
      density profiles before (thin solid line) and after (thick solid
      line) adiabatic compression.  The thin dashed line shows the assumed
      initial baryonic density profile. }
    
    \label{m32Profile}
  \end{center}
\end{figure}

With its extreme central density of 10$^7$ M$_\odot$ pc$^{-3}$ and
two-body core relaxation time of 2--3 Gyr, M32 could possess an
enhanced DM cusp produced through the compression of its DM density
profile by baryonic infall and the growth of the central black hole.
However, it is likely that M32 has undergone multiple mergers during
its evolutionary history which could have significantly altered the DM
density profile.  The degree to which these effects may deplete the
central density of DM in this system is uncertain, and therefore the
estimate of $J$ obtained with the adiabatic compression model should
be considered an upper bound.  We constructed an adiabatic compression
model using the same method applied in the case of M15.  The initial
and final DM and baryon mass profiles for the case of
$m_{vir}=4\times10^{11}$ M$_\odot$ and $c$ = 13.7 are shown in Figure
\ref{m32Profile}. The range for $J$ for the NFW and NFW+AC models is
presented in Table \ref{jtable}.

\subsection{M33}

\begin{figure*}[t]
  \begin{center}
    \includegraphics[width=0.49\textwidth]{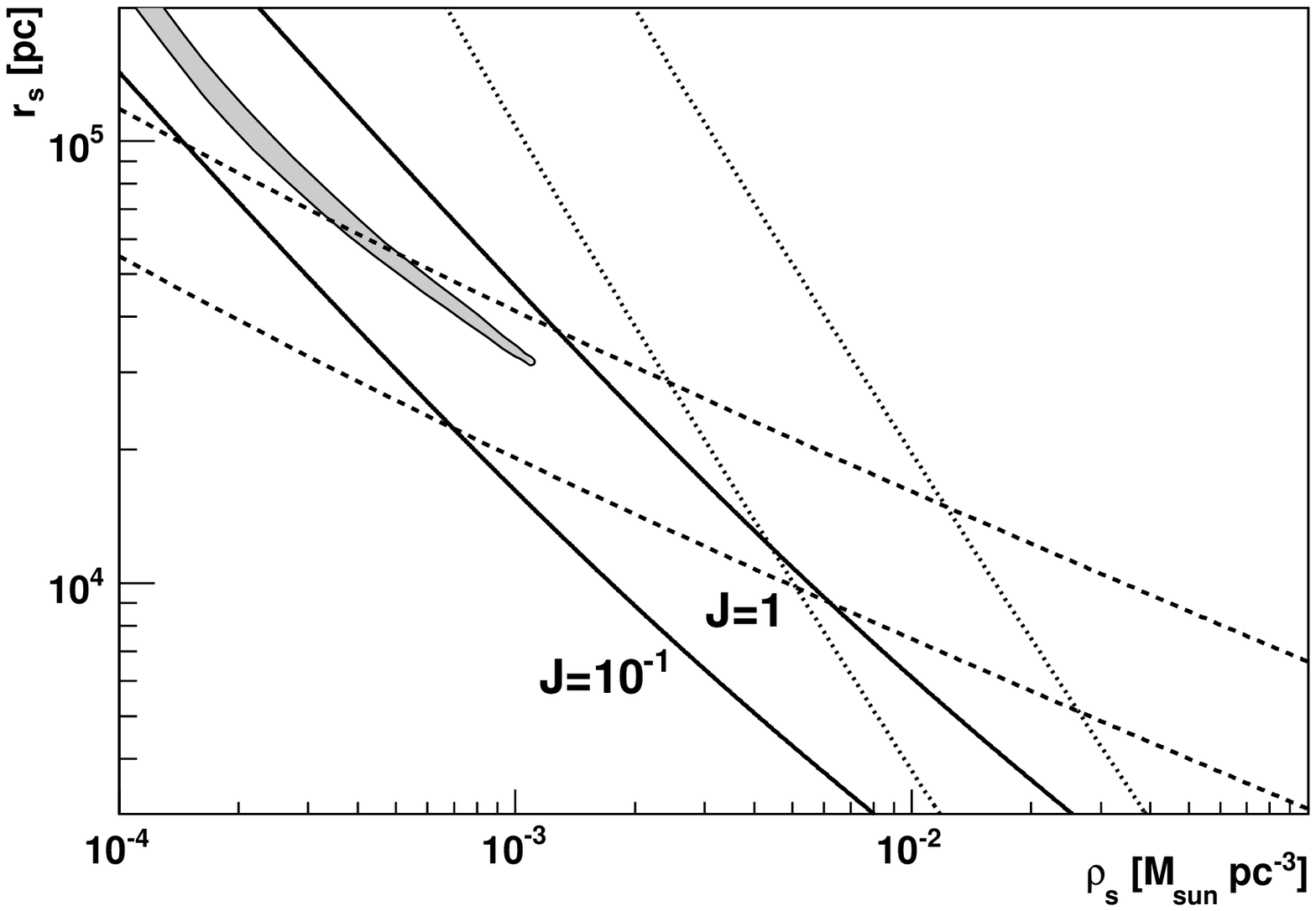}
    \includegraphics[width=0.49\textwidth]{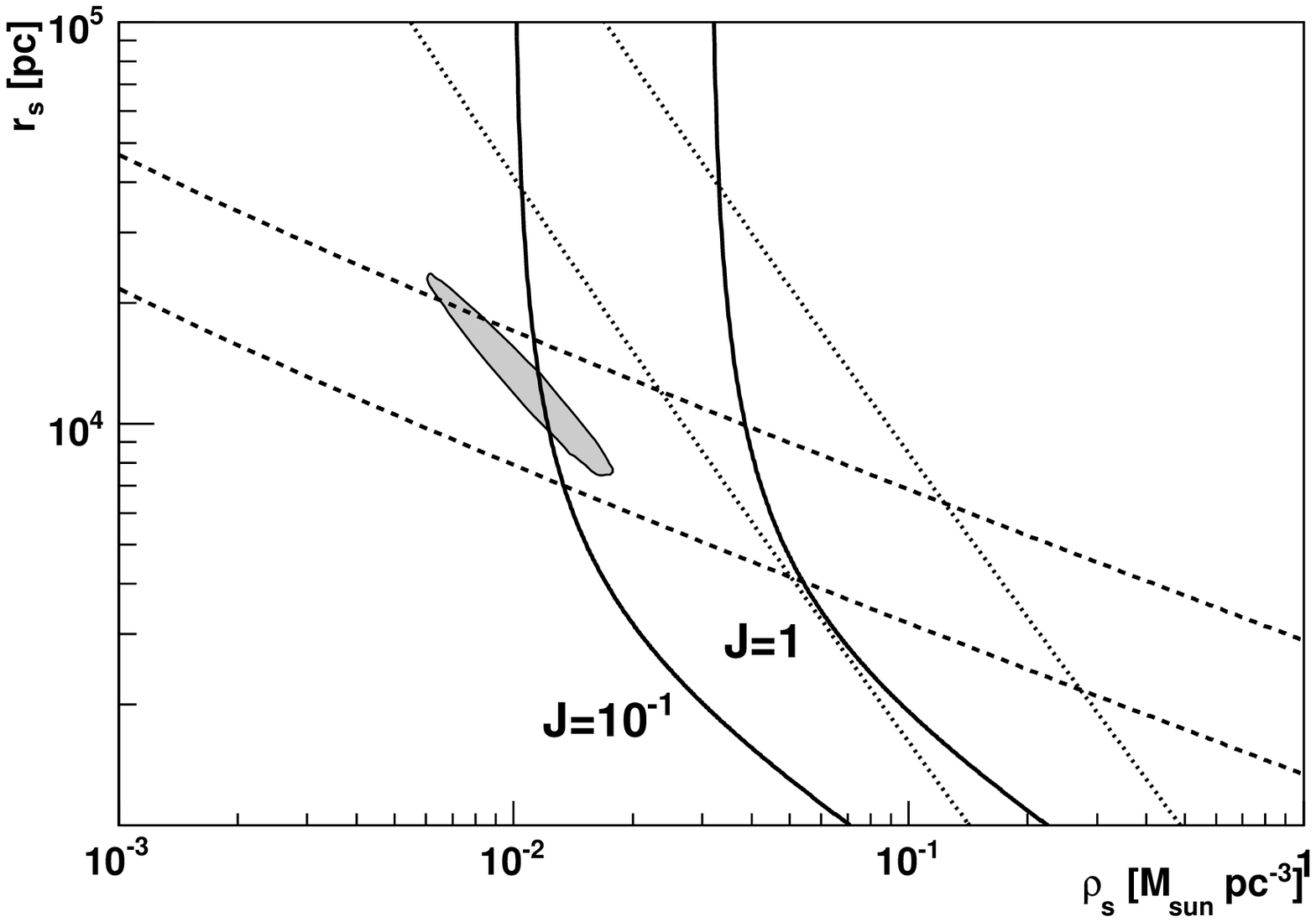}
    
    \caption{Constraints from CO rotation curve data for M33 on the
      parameters $r_{s}$ and $\rho_{s}$ of its DM halo under the
      assumption of an NFW (left panel) and Burkert (right panel)
      profile.  The gray-shaded region indicates the 3 standard
      deviation constraints reported by \citet{corbelli2003}.  Dashed
      lines show contours of constant virial mass for 10$^{11}$
      M$_\odot$ (lower) and 10$^{12}$ M$_\odot$ (upper).  Dotted lines
      denote the $\pm$1 standard deviation bounds on the median halo
      concentration to virial mass relation, Equation
      \ref{concentrationEqn}.  Thick solid lines indicate contours of
      constant $J$.}
    
    \label{m33JPlot}
  \end{center}
\end{figure*}

M33 is a late-type Sc galaxy with a substantial fraction of its
baryonic mass in the form of neutral hydrogen.  Measurements of its
rotation curve imply a lower limit for the mass of its dark halo of
5$\times$10$^{10}$ M$_\odot$ \citep{corbelli2000}.
\citet{corbelli2003} have modeled the rotation curve, as derived from
high-resolution velocity maps of CO, using a three-component density
profile which includes a stellar nucleus of $\sim$8$\times$10$^{8}$
M$_\odot$ interior to $\sim$0.5 kpc, an exponential disk of mass
$\sim$3$\times$10$^{9}$ M$_\odot$, and an extended DM halo with a
scale radius $\sim$20 kpc.  The data is inconsistent with a DM profile
with an intermediate power-law asymptotic as steep as r$^{-1.5}$ as
proposed by \citet{moore1999b} but can be well matched by either a
Burkert or NFW profile.  The best-fit region in the ($\rho_{s},r_{s}$)
plane for both NFW and Burkert profiles is shown in Figure \ref{m33JPlot}.
Ranges for $J$ derived from the mass models of \citet{corbelli2003}
are presented in Table \ref{jtable}.

Observations of the core of M33 have shown evidence for a stellar
nucleus with an effective radius of $\sim$3 pc
\citep{lauer1998,stephens2002} and a central velocity dispersion of
$\sim$24 km s$^{-1}$ \citep{gebhardt2001}.  The exceptionally low
velocity dispersion in this region sets an upper limit on the mass of
a central black hole at $\sim$1.5 $\times 10^{3}$ M$_\odot$ and is
compatible with a relaxation timescale of $\sim$$10^{6}$ yr at 0.1 pc.
Due to the unusually rapid evolutionary timescale in the stellar
nucleus, M33 may be able to sustain a steep DM cusp in its core as DM
could be rapidly replenished after a merger or accretion event.  The
kinematics of this region would be conducive to the growth of an
intermediate mass black hole with a relatively large initial to final
BH mass ratio and the creation of a DM cusp in the vicinity of the BH.
The nucleus of M33 hosts the most luminous steady X-ray source in the
local group which is also associated with a radio source and similar
to the galactic microquasar GRS 1915+105 \citep{dubus2002}.  A small
but significant time variability of 10\% in the X-ray luminosity and
associated variability in spectral shape have also been observed
\citep{parola2003}.  These data could be interpreting as supporting
the existence of an accreting BH in the nucleus of M33.  However, the
exact enhancement to $J$ through these processes is difficult to
estimate given the unknown merger history and therefore an upper bound
for $J$ is not presented.

\begin{deluxetable}{lcccc}
  \tablecolumns{5}
  \tablewidth{0pt}
  \tablecaption{A summary of estimates of $J$ for each source \label{jtable}}
  \tablehead{
    \colhead{Source}&
    \colhead{Distance (kpc)}&
    \colhead{Model}&
    \colhead{$J_{min}$}&
    \colhead{$J_{max}$}
  }
  \startdata
  Draco&80&Burkert&1&10\\
  &&NFW&4&40\\
  \tableline
  Ursa Minor&66&NFW&4&20\\
  \tableline
  M15&10&NFW&7&150\\
  &&NFW+AC&8$\times$10$^{3}$&2$\times$10$^{4}$\\
  \tableline
  M32&776&Burkert&0.6&9\\
  &&NFW&1&9\\
  &&NFW+AC&2$\times$10$^{5}$&10$^{6}$\\
  \tableline
  M33&840&Burkert&0.03&0.2\\
  &&NFW&0.2&0.6\\
  \enddata
  \tablecomments{The contribution of DM substructure may potentially
    enhance the estimates of $J$ shown here by a factor $\gtrsim 2$
    and $\lesssim 100$ as discussed in Section \ref{dmProfileSection}.}
\end{deluxetable}

\section{Limits on SUSY Parameter Space}\label{susyLimitsSection}

\begin{figure}[t]
  \begin{center}

    \includegraphics[width=0.49\textwidth]{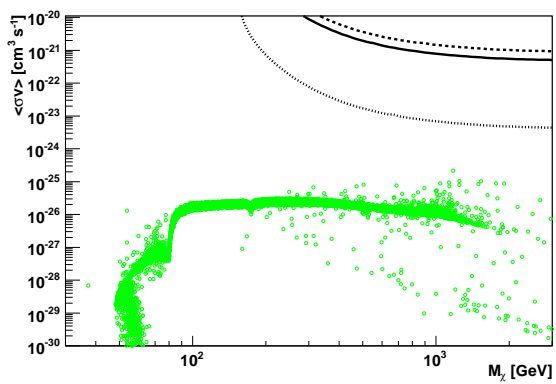}
    \caption{Upper limits on $\left<{\sigma}v\right>$ as a function of
      m$_\chi$ calculated using Equation \ref{diffRateEqn} with a
      composite neutralino spectrum and the $J$ values obtained for
      the Draco (solid line) and Ursa Minor (dashed line) NFW models
      and the M15 NFW+AC model (dotted line).  Shown as open circles
      are MSSM models that fall within $\pm$3 standard deviations of
      the relic density measured in the three-year WMAP data set
      \citep{spergel2007}.}
    \label{mssmLimits}

  \end{center}
\end{figure}

\begin{deluxetable}{lccc}
  \tablecolumns{4}
  \tablewidth{0pt}
  \tablecaption{Upper limits on $\left<{\sigma}v\right>$ of the neutralino 
    \label{crossSectionLimits}}
  \tablehead{
    &
    &
    \multicolumn{2}{l}{95\% C.L. Upper Limit on 
      $\left<{\sigma}v\right>$ for $m_\chi = 1$ TeV (cm$^{3}$ s$^{-1}$)}\\
    \colhead{Source}&
    \colhead{$J$}&
    \colhead{$b\bar{b}$}&
    \colhead{$\tau^{+}\tau^{-}$}
  }
  \startdata
  Draco&13&$<1.9\times 10^{-21}$&$<1.2\times 10^{-22}$\\
  Ursa Minor&9&$<3.9\times 10^{-21}$&$<2.5\times 10^{-22}$\\
  M32&3&$<1.2\times 10^{-20}$&$<8.0\times 10^{-22}$\\
  M33&0.4&$<8.0\times 10^{-20}$&$<5.2\times 10^{-21}$\\
  M15&32&$<5.1\times 10^{-21}$&$<3.3\times 10^{-22}$\\
  \enddata
\end{deluxetable}

For the prediction of gamma-ray fluxes from neutralino
self-annihilation, the framework of the minimal supersymmetric
extension to the standard model (MSSM) was used.  With several
simplifying assumptions, the MSSM can be reduced to the seven
parameters $\mu$, $M_{2}$, tan$\beta$, $m_{A}$, $m_{q}$, $A_{t}$, and
$A_{b}$.  Random scans of these parameters were performed utilizing
the \texttt{DarkSUSY} code \citep{gondolo2005}, and, for each model,
the mass $m_{\chi}$, $\left<{\sigma}v\right>$, and relic density
$\Omega_{DM}h^{2}$ of the neutralino were calculated.  MSSM models
consistent with the $\pm$3 standard deviation bounds on
$\Omega_{DM}h^{2}$ measured by WMAP \citep{spergel2007} were selected
for comparison with the derived limits on $\left<{\sigma}v\right>$.
Figure \ref{mssmLimits} shows a projection in the
$\left<{\sigma}v\right>$--$m_\chi$ plane of MSSM models that satisfy
both the WMAP constraint and the bounds placed by accelerator
experiments.  The majority of the models with neutralino mass above
100 GeV are concentrated in the band with typical
$\left<{\sigma}v\right> \lesssim$ 3 $\times$ 10$^{-26}$ cm$^{3}$
s$^{-1}$ and extending to $m_{\chi}$ $\sim$ 2 TeV.

For a typical choice of MSSM model parameters, the self-annihilation
will predominantly proceed through some combination of the final
states $b\bar{b}$, $t\bar{t}$, $W^{+}W^{-}$, or $ZZ$.  The gamma-ray
spectra of these channels are similar since they all result from the
decay of neutral pions produced in the hadronization of the
annihilation products.  The $\tau^{+}\tau^{-}$ channel produces a
significantly harder spectrum as discussed in Section
\ref{annihilationSpectrumSection}.  Although a neutralino which
annihilates predominantly to the $\tau^{+}\tau^{-}$ channel would
improve the prospects for detection by an ACT, for the MSSM models
considered here, the branching fraction of this channel is never more
than $\sim$10\%.  Three spectra generated using the \texttt{PYTHIA}
Monte Carlo code were adopted to cover the possible range of $dN(E,m_\chi)/dE$:
the $b\bar{b}$ spectrum, the $\tau^{+}\tau^{-}$ spectrum, and a
composite spectrum with $BR(\chi\chi \rightarrow b\bar{b})=0.9$ and
$BR(\chi\chi \rightarrow \tau^{+}\tau^{-}) = 0.1$.

Following Equation \ref{diffFluxEqn}, the upper limit on
$\left<{\sigma}v\right>$ as a function of $m_\chi$ for a source with
an astrophysical enhancement factor $J$ and an upper limit on the
detected rate of gamma rays $R_\gamma\left(95 \% \
  \textrm{C.L.}\right)$ is given by
\begin{equation}\label{diffRateEqn}
\begin{array}{c}\displaystyle
  \left(\frac{\left<{\sigma}v\right>}{3\times10^{-26} \ \textrm{cm}^{3}
      \ \textrm{s}^{-1}}\right) <
  R_\gamma\left(95 \% \ \textrm{C.L.}\right)
  \left(\frac{m_\chi}{100 \ \textrm{GeV}}\right)^2
  \frac{1.45 \times 10^{4}}{J}\\[3mm]\displaystyle
  \times\left[\phi_{1\%} \int_{0}^\infty A(E)
  \left(\frac{dN(E,m_\chi)/dE}{10^{-2} \ \textrm{GeV}^{-1}}\right)dE\right]^{-1},
\end{array}
\end{equation}
where the assumed form of neutralino annihilation spectrum is
convolved with the energy-dependent effective area of the Whipple 10m
telescope, $A(E)$, shown in Figure \ref{whippleEffArea}.  Table
\ref{crossSectionLimits} presents limits on $\left<{\sigma}v\right>$
derived for a neutralino of mass 1 TeV annihilating through the
$b\bar{b}$ and $\tau^{+}\tau^{-}$ channels.  Figure \ref{mssmLimits}
shows the limits on $\left<{\sigma}v\right>$ as a function of $m_\chi$
for the NFW mass models of Draco and Ursa Minor and the adiabatic
compression model (NFW+AC) of M15.  Because the effective area of the
Whipple 10m telescope rapidly declines below $\sim$400 GeV, the limits
on $\left<{\sigma}v\right>$ are most constraining for neutralino
masses above this energy as discussed in Section
\ref{dataAnalysisSection}.

Using the most conservative estimates for $J$, the limits on
$\left<{\sigma}v\right>$ are 10$^{4}$--10$^{5}$ times greater than the
range predicted for the MSSM models considered in this analysis.  The
DM mass models of Draco and Ursa Minor have the best observational
constraints and the fewest uncertainties with regard to the unknown
influence of baryonic matter and merging history.  The lower limit on
$J$ for these galaxies is relatively insensitive to the assumption of
a cusped versus cored DM density profile.  The astrophysical
contribution to the gamma-ray luminosity could be significantly
enhanced if the effects of substructure or a density profile with an
inner logarithmic slope $> 1$ were considered.  Limits derived from
ACT data on $\left<{\sigma}v\right>$ and the monoenergetic line
component $\left<{\sigma}v\right>_{\gamma\gamma}$ have previously been
reported for observations of the G.C.  \citep{aharonian2006} and M31
\citep{aharonian2003a,lavalle2006}.  The limits reported by the
H.E.S.S. collaboration derived from observations of the G.C. are among
the most constraining with a 99\% C.L. upper limit on
$\left<{\sigma}v\right>$ of 10$^{-24}$--10$^{-23}$.  However these
measurements come with a significant systematic uncertainty as they
depend sensitively on the accurate modeling of the astrophysical
background.  For neutralino masses below 100 GeV, EGRET data have also
been shown to constrain $\left<{\sigma}v\right>$ assuming a range of
models for the distribution of DM in substructures and the
distribution of substructures in the Milky Way halo \citep{pieri2007}.

In order for the neutralino to be detectable by the Whipple 10m
telescope, a significant enhancement of the DM density is required.
Such enhancement may be consistent with the kinematics of M15, M32,
and M33.  Among the scenarios discussed, the adiabatic compression
model for M15 provides the most quantitative estimate for $J$.  In the
potentially extreme DM enhancement scenarios which may exist in M32
and M33, current limits on the gamma-ray rate from these sources
already have the potential to constrain the parameter space of allowed
SUSY models.  If one assumes that DM is composed of neutralinos with
$\left<{\sigma}v\right> \simeq$ 3 $\times$ 10$^{-26}$ cm$^{3}$
s$^{-1}$ and $m_\chi \gtrsim$ 400 GeV, then the limits presented in
this work rule out a value of $J$ for these sources in excess of
$\sim$10$^{6}$.

\section{Conclusions}

We have conducted a search for the gamma-ray signature of neutralino
self-annihilation from five sources: the dwarf spheroidal galaxies
Draco and Ursa Minor, the globular cluster M15, and the local group
galaxies M32 and M33.  Each of these sources was chosen as a favorable
representative of different astrophysical conditions that could
potentially enhance the neutralino density and gamma-ray
self-annihilation flux.  For a generic MSSM model of the neutralino,
the self-annihilation flux is only detectable by the Whipple 10m
telescope if such an enhancement is significant.

A standard analysis of the data revealed no significant excesses, and
upper limits on the gamma-ray flux from each source were derived
relative to the flux of the Crab Nebula.  We have derived limits on
$\left<{\sigma}v\right>$ of the neutralino as a function of its mass
using models for the DM profile of each source and the generic
differential gamma-ray spectrum of the neutralino self-annihilation
constructed from two representative channels, $b\bar{b}$ and
$\tau^{+}\tau^{-}$.  Using the upper limit on the gamma-ray rate
measured from Draco and the most conservative estimate of the DM
distribution in this source, we obtain 95 \% C.L.  upper limits on
$\left<{\sigma}v\right>$ of $<1.9\times10^{-21}$ cm$^{3}$ s$^{-1}$ and
$<1.2\times10^{-22}$ cm$^{3}$ s$^{-1}$ for a neutralino of mass 1 TeV
annihilating exclusively through the $b\bar{b}$ and $\tau^{+}\tau^{-}$
channels, respectively.

We have considered potential enhancements to the DM self-annihilation
flux including the effects of DM substructure and the adiabatic
compression of the DM halo due to baryonic infall.  These scenarios
could enhance the annihilation flux by as much as $10^{4}$ and
possibly higher.  However the uncertainties of these estimates are
large due to the poorly understood dynamics in the cores of the
objects as well as their potentially complex merging histories.  If
one assumes that DM is composed of neutralinos with
$\left<{\sigma}v\right> \simeq 3 \times$ 10$^{-26}$ cm$^{3}$
s$^{-1}$, then some extreme enhancement scenarios for M32 and M33 may
be ruled out.

The current generation of ACTs such as VERITAS, H.E.S.S., MAGIC, and
CANGAROO-III have the potential to improve significantly the
sensitivity of these measurements and thus probe a larger region of
the MSSM parameter space.  For a source with a Crab Nebula-like
spectrum, VERITAS has a flux sensitivity $\sim$10 times better than
the Whipple 10m telescope and a peak detection rate near 150 GeV.  The
lower energy threshold of VERITAS will allow it to be sensitive to the
low to intermediate neutralino mass range of 100 GeV--1 TeV.  With the
improved sensitivity of these instruments, conventional astrophysical
backgrounds may become significant.  The sensitivity of observations
of the G.C. to DM annihilations is already limited by the presence of
such backgrounds.  Observations of extragalactic sources such as those
discussed in this work have the potential to avoid this limitation.
Furthermore, the identification of the unique spectral signature of DM
self-annihilations in two or more of these sources would effectively
rule out a traditional astrophysical process.  Next-generation ACT
instruments such as the currently planned Cherenkov Telescope Array
(CTA) and Advanced Gamma Ray Imaging System (AGIS) will potentially be
10$^{2}$--10$^{3}$ times as sensitive as the Whipple 10m telescope and
could perform dedicated deep observations with 10--10$^{2}$ times
longer exposure than the observations presented in this work.  This
may allow the exclusion of MSSM models with even the most conservative
assumptions for the DM distribution in the sources with the lowest
anticipated astrophysical backgrounds such as the dwarf spheroidal
galaxies.

\acknowledgements
This research is supported by grants from the U.S. National Science
Foundation, the U.S. Department of Energy, the Smithsonian Institution, by
NSERC in Canada, by Science Foundation Ireland, and by PPARC in the UK.
V.V.V. acknowledges the support of the U.S. National Science Foundation
under CAREER program (Grant No. 0422093).

\end{document}